\documentstyle[epsfig,a4,12pt,rotating,here,amssymb,enumerate,cite,here]{article}
\begin{document}
\setlength{\leftmargin}{-5cm}
\setlength{\topmargin}{-1.0cm}
\setlength{\textwidth}{16cm}
\setlength{\textheight}{24.5cm}
\newcommand{\PNnum}     {OPAL PAPER PR XXX}
\newcommand{\Date}      {\today}
\newcommand{\Author}    {The OPAL Higgs group}
\def\mrm{\mathrm}
\def\mbf{\mathbf}
\newcommand {\genH}      {\mathrm{{\cal H}}}
\newcommand {\genmH}      {\mathrm{m_{\cal H}}}
\newcommand {\genHone}      {\mathrm{{{\cal H}_1}}}
\newcommand {\genHtwo}      {\mathrm{{{\cal H}_2}}}
\newcommand {\genmHone}      {\mathrm{m_{{\cal H}_1}}}
\newcommand {\genmHtwo}      {\mathrm{m_{{\cal H}_2}}}
\newcommand{\cls}        {\mbox{$\mathrm{CL_s}$}}
\newcommand{\clb}        {\mbox{$\mathrm{CL_b}$}}
\newcommand{\ecm}        {\mbox{$E_{\mathrm CM}$}}
\newcommand{\sprime}     {\mbox{$\sqrt{s^\prime}$}}
\newcommand{\eerzrh}{\mbox{$\mathrm{e}^+\mathrm{e}^-\rightarrow\mathrm{Z}^{0}\rightarrow\mathrm{hadrons}$}}
\newcommand{\eerhz}{\mbox{$\mathrm{e}^+\mathrm{e}^-\rightarrow\mathrm{h}^{0}\mathrm{Z}^{0}$}}
\newcommand{\erzgrqqg}{\mbox{$\mathrm{e}^+\mathrm{e}^-\rightarrow\mathrm{Z^0\gamma}\rightarrow\mathrm{q\bar{q}\gamma}$}}
\newcommand{\zgsrqq}{\mbox{$\mathrm{(Z/\gamma)}^{*}\rightarrow\mathrm{q\bar{q}}$}}
\newcommand{\eerqqqq}{\mbox{$\mathrm{e}^+\mathrm{e}^-\rightarrow\mathrm{q\bar{q}q\bar{q}}$ and $\mathrm{q\bar{q}}gg$}}
\newcommand{\zzqqll}{\mbox{$\Z\Z^{(*)}\ra\qq\ellell$}}
\newcommand{\glgl}{\mbox{$\mathrm{g}\mathrm{g}$}} 
\newcommand{\mHsm}{\mbox{$m_{\mathrm{H}^0_{SM}}$}}
\newcommand{\Hsm}{\mbox{$\mathrm{H}^{0}_{SM}$}}
\newcommand{\Z}{\mbox{$\mathrm{Z}^{0}$}}
\newcommand{\A}{\mbox{$\mathrm{A}^{0}$}}
\newcommand{\sqrtsp}{\mbox{$\sqrt{s'}$}}
\newcommand{\epm} {\mbox{$\mathrm{e}^+ \mathrm{e}^-$}}
\newcommand{\mpm} {\mbox{$\mu^+ \mu^-$}}
\newcommand{\nprode}{\mbox{$N^{\mathrm{e}^+ \mathrm{e}^-}_{prod}$}}
\newcommand{\nprodm}{\mbox{$N^{\mu^+ \mu^-}_{prod}$}}
\newcommand{\nexpe}{\mbox{$N_{exp}^{\mathrm{e}^+ \mathrm{e}^-}$}}
\newcommand{\nexpm}{\mbox{$N_{exp}^{\mu^+ \mu^-}$}}
\newcommand{\nexpt}{\mbox{$N_{exp}^{total}$}}
\newcommand{\bZo}{{\bf \mbox{$\mathrm Z}^0$}}
\newcommand{\Hosm}{\mbox{$\mathrm{H}^{0}_{\mathrm{SM}}$}}
\newcommand{\Wp}{\mbox{$\mathrm{W}^+$}}
\newcommand{\Wm}{\mbox{$\mathrm{W}^-$}}
\newcommand{\Hpm}{\mbox{$\mathrm{H}^{\pm}$}}
\newcommand{\Hp}{\mbox{$\mathrm{H}^+$}}
\newcommand{\Hm}{\mbox{$\mathrm{H}^-$}}
\newcommand {\HH}{\Hp\Hm}
\newcommand{\ZZ}{\mbox{$\mathrm{Z}^{0}{\mathrm{Z}^{0}}^{(*)}$}}
\newcommand{\ZZz}{\mbox{$\mathrm{Z}^{0}{\mathrm{Z}^{0}}$}}
\newcommand{\sctop}{\mbox{$\tilde{\mathrm{t}}$}}
\newcommand{\msctop}{\mbox{$m_{\tilde{\mathrm{t}}}$}}
\newcommand{\ko}{\mbox{${\tilde{\chi_1}^0}$}}
\newcommand{\koko}{\mbox{${\tilde{\chi_1}^0}{{\tilde{\chi_1}^0}}$}}
\newcommand{\bho}{\mbox{$\boldmath{\mathrm{H}^{0}}$}}
\newcommand{\leplep}{\mbox{$\ell^+\ell^-$}}
\newcommand{\bee}{\mbox{$\boldmath {\mathrm{e}^{+}\mathrm{e}^{-}} $}}
\newcommand{\bmm}{\mbox{$\boldmath {\mu^{+}\mu^{-}} $}}
\newcommand{\bnn}{\mbox{$\boldmath {\nu \bar{\nu}} $}}
\newcommand{\bqq}{\mbox{$\boldmath {\mathrm{q} \bar{\mathrm{q}}} $}}
\newcommand{\qqp}{\mbox{$\mathrm{q\overline{q}^\prime}$}}
\newcommand{\qpq}{\mbox{$\mathrm{q^\prime\overline{q}}$}}
\newcommand{\qppqppp}{
            \mbox{$\mathrm{q^{\prime\prime}\overline{q}^{\prime\prime\prime}}$}}
\newcommand{\tnt}{\mbox{${\tau\nu_{\tau}}$}}
\newcommand{\tpnu}{\mbox{${\tau^+\nu_{\tau}}$}}
\newcommand{\tmnu}{\mbox{${\tau^-{\bar{\nu}}_{\tau}}$}}
\newcommand{\lnu}{\mbox{$\ell\nu$}}
\newcommand{\tnu}{\mbox{$\tau\nu$}}
\newcommand{\br}{\mbox{$\boldmath {\rightarrow}$}}
\newcommand{\erh}{\mbox{$\mathrm{e}^+\mathrm{e}^-\rightarrow\mathrm{hadrons}$}}
\newcommand{\tpm}{\mbox{$\tau^{\pm}$}}
\newcommand{\pzvis}{\mbox{$\protect P^z_{\mathrm vis}$}}
\newcommand{\evisn}{\mbox{$\protect E_{\mathrm vis}$/$\protect \sqrt{s}$}}
\newcommand{\rvis}{\mbox{$R_{\mrm{vis}}$}}
\newcommand{\wwln}{\mbox{$\mrm{W}^{+}\mrm{W}^{-}
            \ra\mathrm{q}\bar{{\mathrm{q}}^{\prime}}\lnu$}}
\newcommand{\gamgam}{\mbox{$\gamma\gamma$}}
\newcommand{\uu}{\mbox{$\mathrm{u} \bar{\mathrm{u}}$}}
\newcommand{\dd}{\mbox{$\mathrm{d} \bar{\mathrm{d}}$}}
\newcommand{\cc}{\mbox{$\mathrm{c} \bar{\mathrm{c}}$}}
\newcommand{\csbar}{\mbox{$\mathrm{c} \bar{\mathrm{s}}$}}
\newcommand{\cbars}{\mbox{$\bar{\mathrm{c}} \mathrm{s}$}}
\newcommand{\mhs}{\mbox{$m^2_{\mathrm{h}^{0}}$}}
\newcommand{\mHpm}{\mbox{$m_{\mathrm{h}^{\pm}}$}}
\newcommand{\mHp}{\mbox{$m_{\mathrm{H}^+}$}}
\newcommand{\mHm}{\mbox{$m_{\mathrm{H}^-}$}}
\newcommand{\mW}{\mbox{$m_{\mathrm{W}^{\pm}}$}}
\newcommand{\mWn}{\mbox{$m_{\mathrm{W}}$}}
\newcommand{\mtop}{\mbox{$m_{\mathrm{t}}$}}
\newcommand{\mstop}{\mbox{$m_{\tilde{\mathrm{t}}}$}}
\newcommand{\mstopL}{\mbox{$m_{\tilde{\mathrm{t}}_{\mathrm{L}}}$}}
\newcommand{\mstopR}{\mbox{$m_{\tilde{\mathrm{t}}_{\mathrm{R}}}$}}
\newcommand{\mb}{\mbox{$m_{\mathrm{b}}$}}
\newcommand{\lpm}{\mbox{$\ell ^+ \ell^-$}}
\newcommand{\G}{\mbox{$\mathrm{GeV}$}}
\newcommand{\Gc}{\mbox{$\mathrm{GeV}$}}
\newcommand{\Gcs}{\mbox{$\mathrm{GeV}$}}
\newcommand{\epsnn}{\mbox{$\epsilon^{\nu\bar{\nu}}$(\%)}}
\newcommand{\Nnn}{\mbox{$N^{\nu \bar{\nu}}_{exp}$}}
\newcommand{\epsll}{\mbox{$\epsilon^{\ell^{+}\ell^{-}}$(\%)}}
\newcommand{\Nll}{\mbox{$N^{\ell^+\ell^-}_{exp}$}}
\newcommand{\Nexp}{\mbox{$N^{total}_{exp}$}}
\newcommand{\kl}{\mbox{$\mathrm{K_{L}}$}}
\newcommand{\dedx}{\mbox{d$E$/d$x$}}
\newcommand{\ie}{\mbox{$i.e.$}}
\newcommand{\sba}{\mbox{$\sin ^2 (\beta -\alpha)$}}
\newcommand{\cba}{\mbox{$\cos ^2 (\beta -\alpha)$}}
\newcommand{\Dbi}{\mbox{${\cal B}_i$}}
\newcommand{\Dbj}{\mbox{${\cal B}_{\mathrm{jet}}$}}
\newcommand{\ee}{\mbox{${\mathrm{e}}^+ {\mathrm{e}}^-$}}
\newcommand{\tautau}{\mbox{$\tau^+\tau^-$}}
\newcommand{\mm}{\mbox{$\mu^+\mu^-$}}
\newcommand{\ellell}{\mbox{$\ell^+\ell^-$}}
\newcommand{\qq}         {\mbox{$\mathrm{q}\bar{\mathrm{q}}$}}
\newcommand{\bb}         {\mbox{$\mathrm{b}\bar{\mathrm{b}}$}}
\newcommand{\ff}         {\mbox{$\mathrm{f}\bar{\mathrm{f}}$}}
\newcommand{\nunu}       {\mbox{$\nu\bar{\nu}$}}
\newcommand{\mZ}         {\mbox{$m_{\mathrm{Z}}$}}
\newcommand{\mH}         {\mbox{$m_{\mathrm{H}}$}}
\newcommand{\mh}         {\mbox{$m_{\mathrm{h}}$}}
\newcommand{\mA}         {\mbox{$m_{\mathrm{A}}$}}
\newcommand{\czcz}{\mbox{$\chi^0_1\chi^0_1$}}
\newcommand{\cz}{\mbox{$\chi^{0}$}}
\newcommand{\co}{\mbox{${\tilde{\chi}_1^0}$}}
\newcommand{\ct}{\mbox{${\tilde{\chi}_2^0}$}}
\newcommand{\coct}{\mbox{$\chi^0_1\chi^0_2$}}
\newcommand{\ctcoz}{\ct\ra\Zs}
\newcommand{\ctcog}{$\ct\ra\co\gamma$}
\newcommand{\hczcz}{\ho\ra\co\co}
\newcommand{\hcoct}{\ho\ra\co\ct}
\newcommand {\Ho}        {\mbox{$\mathrm{H}^{0}$}}
\newcommand {\Ao}        {\mbox{$\mathrm{A}^{0}$}}
\newcommand {\ho}        {\mbox{$\mathrm{h}^{0}$}}
\newcommand {\Zo}        {\mbox{$\mathrm{Z}^{0}$}}
\newcommand{\Zs}         {\mbox{${\mathrm{Z}}^{*}$}}
\newcommand{\Zgs}        {\mbox{$\mathrm{(Z/\gamma)}^{*}$}}
\newcommand {\Wpm}       {\mbox{$\mathrm{W}^{\pm}$}}
\newcommand{\MH}{M_{\mathrm H}}
\newcommand{\MZ}{M_{\mathrm Z}}
\newcommand{\qqbar}{{\mathrm q}\bar{\mathrm q}}
\newcommand{\nn}{\mbox{$\nu \bar{\nu}$}}
\newcommand{\gaga}       {\mbox{$\gamma\gamma$}}
\newcommand{\WW}         {\mbox{$\mathrm{W}^+\mathrm{W}^-$}}
\newcommand{\pb}         {\mbox{$\mathrm{pb}^{-1}$}}
\newcommand{\tanb}       {\mbox{$\tan\!\beta$}}
\newcommand{\h}{\mbox{$\mathrm{h}^{0}$}}
\def\mrm       {\mathrm}
\newcommand{\gsim}{\;\raisebox{-0.9ex}
           {$\textstyle\stackrel{\textstyle >}{\sim}$}\;}
\newcommand{\lsim}{\;\raisebox{-0.9ex}{$\textstyle\stackrel{\textstyle<}
           {\sim}$}\;}
\newcommand{\degree}    {^\circ}
\newcommand{\sqrts}     {\mbox{$\sqrt{s}$}}
\newcommand{\ssq}     {\mbox{$\sqrt{s}$}}
\newcommand{\PhysLett}  {Phys.~Lett.}
\newcommand{\PRL}       {Phys.~Rev.\ Lett.}
\newcommand{\PhysRep}   {Phys.~Rep.}
\newcommand{\PhysRev}   {Phys.~Rev.}
\newcommand{\NPhys}     {Nucl.~Phys.}
\newcommand{\ZPhysC}[1]    {Z. Phys. {\bf C#1}}
\newcommand{\EurPhysC}[1]    {Eur. Phys. Journal {\bf C#1}}
\newcommand{\PhysLettB}[1] {Phys. Lett. {\bf B#1}}
\newcommand{\CPC}[1]      {Comp.\ Phys.\ Comm.\ {\bf #1}}
\def\etal{\mbox{{\it et al.}}}
\newcommand{\NIM} {Nucl.~Instr.\ Meth.}
\newcommand{\ZPhys}  {Z.~Phys.}
\newcommand{\IEEENS} {IEEE Trans.\ Nucl.~Sci.}
\newcommand{\OPALColl}  {OPAL Collab.}
\newcommand{\ra}        {\mbox{$\rightarrow$}}   
\newcommand{\fcf}{$5\frac{1}{2}$~C fit}
\newcommand{\me}{matrix element}

\begin{titlepage}
\begin{center}
{\large{EUROPEAN ORGANIZATION FOR NUCLEAR RESEARCH}}
\end{center}
\bigskip
\begin{flushright}
      CERN-PH-EP/2004-039 \\
      27th July, 2004 \\
\end{flushright}
\bigskip
\bigskip

\begin{center}{\Large\bf  
Flavour Independent ${\mathrm\mathbf{h^0A^0}}$ Search and 
Two Higgs Doublet Model Interpretation 
                          of Neutral Higgs Boson Searches at LEP}\\
\bigskip\bigskip
{\Large{The OPAL Collaboration}}
\bigskip\bigskip
\end{center}
\bigskip
\begin{center}{\large  Abstract}\end{center}

\hspace{-0.75cm}

Upper limits on the cross-section of the pair-production process \ee\ra\ho\Ao, 
assuming 100 $\%$ decays into hadrons,
are derived from a new search for the \ho\Ao\ra\ hadrons topology,
independent of the hadronic flavour of the decay products.
Searches for the neutral Higgs bosons \ho\ and \Ao,
are used to obtain constraints on 
the Type II Two Higgs Doublet Model (2HDM(II)) with 
no CP violation in the Higgs sector and no additional non Standard Model
particles besides the five Higgs bosons.
The analysis combines LEP1 and LEP2 
data collected with the OPAL detector 
up to the highest available centre-of-mass energies.
The searches are sensitive to the
\ho, \Ao \ra\qq, \glgl, \tautau\ and \ho\ra\Ao\Ao\ decay modes 
of the Higgs bosons. 
The 2HDM(II) parameter space is explored 
in a detailed scan. 
Large regions of the 2HDM(II) parameter space
are excluded at the 95\% CL in the (\mh, \mA),
(\mh, \tanb) and (\mA, \tanb) planes,
using both direct neutral Higgs boson searches and indirect
limits derived from Standard Model high precision measurements.
The region 1 $\lesssim$ \mh $\lesssim$ 55 GeV and
3 $\lesssim$ \mA $\lesssim$ 63 GeV is excluded at 95~\% CL
independent of the choice of the 2HDM(II) parameters.



\bigskip\bigskip\bigskip\bigskip\bigskip\bigskip\bigskip\bigskip
\bigskip\bigskip\bigskip\bigskip\bigskip\bigskip\bigskip\bigskip
\bigskip\bigskip\bigskip
\begin{center}
{\large{(Submitted to European Pysical Journal C)}}
\end{center}
\end{titlepage}

\begin{center}{\Large        The OPAL Collaboration
}\end{center}\bigskip
\begin{center}{
G.\thinspace Abbiendi$^{  2}$,
C.\thinspace Ainsley$^{  5}$,
P.F.\thinspace {\AA}kesson$^{  3,  y}$,
G.\thinspace Alexander$^{ 22}$,
J.\thinspace Allison$^{ 16}$,
P.\thinspace Amaral$^{  9}$, 
G.\thinspace Anagnostou$^{  1}$,
K.J.\thinspace Anderson$^{  9}$,
S.\thinspace Asai$^{ 23}$,
D.\thinspace Axen$^{ 27}$
% ...
}\end{center}\bigskip
\bigskip
$^{  1}$School of Physics and Astronomy, University of Birmingham,
Birmingham B15 2TT, UK
\newline
$^{  2}$Dipartimento di Fisica dell' Universit\`a di Bologna and INFN,
I-40126 Bologna, Italy
\newline
$^{  3}$Physikalisches Institut, Universit\"at Bonn,
D-53115 Bonn, Germany
\newline
% ...
$^{  5}$Cavendish Laboratory, Cambridge CB3 0HE, UK
\newline
% ...
$^{  9}$Enrico Fermi Institute and Department of Physics,
University of Chicago, Chicago IL 60637, USA
\newline
% ...
$^{ 16}$Department of Physics, Schuster Laboratory, The University,
Manchester M13 9PL, UK
\newline
% ...
$^{ 22}$Department of Physics and Astronomy, Tel Aviv University,
Tel Aviv 69978, Israel
\newline
$^{ 23}$International Centre for Elementary Particle Physics and
Department of Physics, University of Tokyo, Tokyo 113-0033, and
Kobe University, Kobe 657-8501, Japan
\newline
% ...
$^{ 27}$University of British Columbia, Department of Physics,
Vancouver BC V6T 1Z1, Canada
\newline
% ...
$^{  b}$ now at University of Iowa, Dept of Physics and Astronomy, Iowa, U.S.A. 
\newline
$^{  c}$ and Institute of Nuclear Research, Debrecen, Hungary
\newline
$^{  e}$ and Department of Experimental Physics, University of Debrecen, 
Hungary
\newline
$^{  f}$ and MPI M\"unchen
\newline
$^{  g}$ and Research Institute for Particle and Nuclear Physics,
Budapest, Hungary
\newline
$^{  h}$ now at University of Liverpool, Dept of Physics,
Liverpool L69 3BX, U.K.
\newline
$^{  i}$ now at Dept. Physics, University of Illinois at Urbana-Champaign, 
U.S.A.
\newline
% ...
$^{  q}$ now at IPHE Universit\'e de Lausanne, CH-1015 Lausanne, Switzerland
\newline
% ...
$^{  s}$ now at University of Antwerpen, Physics Department,B-2610 Antwerpen, 
Belgium; supported by Interuniversity Attraction Poles Programme -- Belgian
Science Policy
\newline
$^{  u}$ and High Energy Accelerator Research Organisation (KEK), Tsukuba,
Ibaraki, Japan
\newline
% ...
$^{  x}$ now at DESY Zeuthen
\newline
$^{  y}$ now at CERN
\newline
% ...

\newpage
\section{Introduction}
\label{sect:intro}

The data collected by the OPAL detector at LEP during the years 1999 and 2000
at centre-of-mass energies $\sqrt{s} ~\approx~ 192,~ 196,~ 200-209$~GeV are combined 
with the data at the \Zo\ pole, $\sqrts~\approx~ 183$ GeV and 189 GeV,
to search  for 
neutral Higgs bosons \cite{higgs, higgsEng, higgsGur} 
in the framework of the Type II Two Higgs Doublet Model (2HDM(II)) \cite{hollik, hollik2}  
with no CP violation in the Higgs sector 
and no additional particles besides those arising from the Higgs mechanism.
This study updates the results of a previous 
OPAL publication  \cite{2hdmpaper}, which included data at 
the \Zo\ pole, $\sqrts~\approx~183$ GeV and 189 GeV.

In the minimal Standard Model (SM) the Higgs sector 
comprises only one complex Higgs doublet \cite{higgs}
resulting in one physical neutral Higgs 
scalar whose mass is a free parameter of the theory. 
However, it is important to study extended models  
containing more than one physical Higgs boson 
in the spectrum. 
In particular, Two Higgs Doublet Models (2HDMs) are attractive extensions of the SM
since they add new phenomena with 
the fewest new parameters; they satisfy the 
constraints of $\rho \approx 1$~\cite{higgshunter} and the absence 
of tree-level flavour changing neutral currents,
if the Higgs-fermion couplings are appropriately chosen.

In the context of 2HDMs
the Higgs sector comprises five physical Higgs bosons: 
two neutral CP-even scalars, \ho\ and \Ho\ (with  $\mh < \mH$), one
CP-odd scalar, \Ao, and two charged scalars, \Hpm. 
The four Higgs masses are free parameters of the model.

Within 2HDMs the choice of the couplings between the Higgs bosons and
the fermions determines the type of the model considered. In the Type II 
model the first Higgs doublet 
couples only to down--type fermions and the 
second Higgs doublet couples only to up--type 
fermions. In the Type I model the 
quarks and leptons only couple to the second Higgs 
doublet.
The Higgs sector in the 
minimal supersymmetric extension of the SM \cite{higgshunter,fayet}
is a Type II 2HDM, in which the introduction of 
supersymmetry adds new particles and constrains the parameter space of
the Higgs sector of the model. 
The 2HDM(II) Higgs potential and detailed description of
the physical parameters of the model are  
given in \cite{hollik, hollik2,higgshunter}. 

At the centre-of-mass energies accessed by
LEP, the \ho\ and \Ao\  
bosons are expected to be produced predominantly via two processes: 
the {\it{Higgs-strahlung}}
process \ee\ra\ho\Zo\ 
and the {\it{pair--production}} process \ee\ra\ho\Ao.
The cross-sections for these two processes,
$\sigma_{\mathrm{hZ}}$ and $\sigma_{\mathrm{hA}}$,
are related at tree-level 
to the SM Higgs-strahlung production cross-section by the following relations \cite{higgshunter}: 
\begin{eqnarray}
\ee\ra\ho\Zo\;:&&
\sigma_{\mathrm{hZ}}~=~\sin^2(\beta -\alpha)~\sigma^{\mathrm{SM}}_{\mathrm{HZ}},
\label{equation:xsec_zh} \\
\ee\ra\ho\Ao\;:&&
\sigma_{\mathrm{hA}}~=~
\cos^2(\beta-\alpha)~\bar{\lambda}~\sigma^{\mathrm{SM}}_{\mathrm{HZ}},
\label{equation:xsec_ah}
\end{eqnarray} 
where $\sigma^{\mathrm{SM}}_{\mathrm{HZ}}$ is the Higgs-strahlung cross-section 
for the SM process \ee\ra\Hosm\Zo, 
$\alpha$ is the Higgs mixing angle, 
\tanb\ is defined in terms of the ratio of the
vacuum expectation values, $v_1$ and $v_2$, of the two scalar fields,
$\tanb = v_2/v_1$ and
$\bar{\lambda} = \lambda^{3/2}_{\rm{Ah}} /\{{\lambda^{1/2}_{\rm{Zh}}
[12 m_{\rm{Z}}^2/s + \lambda_{\rm{Zh}}]}\} $
accounts for the suppression of the P-wave cross-section near the 
threshold, with 
$\lambda_{ij} = (1- m_i^2/s + m_j^2/s)^2 - 4 m_i^2m_j^2/s^2$
being the two--particle phase--space factor.

In a 2HDM the production cross-sections and Higgs boson decay branching ratios
are predicted for a given set of model parameters.
The coefficients \sba\ and \cba\ which appear
in Eqs.~(\ref{equation:xsec_zh}) and~(\ref{equation:xsec_ah})
determine the production cross-sections.  The decay branching ratios to
the various final states 
are also determined by $\alpha$ and $\beta$.
In the 2HDM(II) the tree-level couplings of the \ho\ and \Ao\ bosons to the up-- and down--type 
quarks relative to the couplings of the SM Higgs boson to the corresponding
fermions are \cite{higgshunter}
\begin{equation}
\label{eq:BRs}
{\mathrm{h^0}} {\mathrm{c}} \overline{{\mathrm{c}}} : \frac{\cos \alpha}{\sin\beta},~~~~
{\mathrm{h^0}} {\mathrm{b}} \overline{{\mathrm{b}}} : -~\frac{\sin \alpha}{\cos\beta},~~~~
{\mathrm{A^0}} {\mathrm{c}} \overline{{\mathrm{c}}} : \cot \beta,~~~~
{\mathrm{A^0}} {\mathrm{b}} \overline{{\mathrm{b}}} : \tan \beta,
\end{equation}
indicating the need for a scan over the range of both angles 
when considering the different production 
cross-section mechanisms and final state topologies.

In the analysis described in this paper, 
detailed scans over broad ranges of these parameters are performed.
Each of the scanned points is considered as an independent scenario
within the 2HDM(II), and results are provided for each point in the
(\mh,~\mA,~\tanb,~$\alpha$) space.
The masses \mh\ and \mA\ are varied such that 
the kinematically accessible range at LEP is fully covered.
The choice  $0 < \beta < \pi/2$ is derived from $v_1,v_2>0$ which
in the MSSM implies that $-\pi/2 \le \alpha \le 0$~\cite{hunters24}. 
This range was studied in \cite{2hdmpaper} to cover
an MSSM oriented 2HDM(II). 
However, to completely cover any 2HDM(II),  
$\alpha$ has to be varied over an arbitrarily chosen
angular range of $\pi$.
In order to extend the analysis done in \cite{2hdmpaper}
beyond the MSSM--like 2HDM(II), 
the domain $-\pi/2 \le \alpha \le \pi/2$ is explored in the present study.
The final-state topologies of the processes (\ref{equation:xsec_zh}) and
(\ref{equation:xsec_ah}) are determined by the decays of the \Z,
\h\ and \A\ bosons. Higgs bosons couple to fermions 
with a strength proportional to the fermion mass, 
favouring the decays into pairs of 
b--quarks and tau leptons at LEP energies. 
However, with values of $\alpha$ and \tanb\ close to zero
the decays into up--type light quarks and gluons through quark loops become dominant,
motivating the inclusion of flavour independent analyses.

Section~\ref{sect:detector} contains 
a short description of the OPAL detector and the Monte Carlo simulations used. 
A new analysis at the highest LEP energies 
with improved sensitivity for the process \ee\ra\ho\Ao\ra\ hadrons, independent of the 
hadronic flavour of the decay products, is presented in
Section~\ref{sect:fiha}. In Section~\ref{sect:combmet}
the confidence level calculation method is described.
A limit on the cross-section of
the pair-production process \ee\ra\ho\Ao, assuming 
100 $\%$ decays into hadrons, is given in Section~\ref{sect:modind}.
The data samples, the final state topologies studied and
the external constraints used for the 2HDM(II) interpretation
are described in Section~\ref{sect:datasamplesall}.
The 2HDM(II) interpretation of the searches
is presented in Section~\ref{section:limits}, and
in Section~\ref{section:conclusion} the results are summarised 
and conclusions are drawn.

\section{OPAL detector and Monte Carlo samples}\label{sect:detector}
The OPAL detector~\cite{detector} has
nearly complete solid angle coverage and excellent hermeticity.
The innermost detector of the central tracking is a high-resolution
silicon microstrip vertex detector~\cite{simvtx} which lies immediately
outside of the beam pipe.  
The silicon microvertex detector
is surrounded by a high precision 
vertex drift chamber,
a large volume jet chamber, and $z$--chambers to measure the $z$ coordinates\footnote{
OPAL uses a right-handed
coordinate system where the $+z$ direction is along the electron beam and
where $+x$ points to the centre of the LEP ring.  
The polar angle, $\theta$, is
defined with respect to the $+z$ direction and the azimuthal angle, $\phi$,
with respect to the horizontal, $+x$ direction.}
of tracks, all in a uniform
0.435~T axial magnetic field. The lead-glass electromagnetic calorimeter
and the presampler are located outside the magnet coil.  It provides,
in combination with a number of forward detectors
and the silicon-tungsten luminometer~\cite{sw}, geometrical acceptance
down to 25~mrad from the beam direction.  The silicon-tungsten luminometer
serves to measure the integrated luminosity using small angle Bhabha
scattering events~\cite{lumino}.
The magnet return yoke is instrumented with streamer tubes and thin gap
chambers for hadron calorimetry and is surrounded by several layers 
of muon chambers.

Events are reconstructed from charged particle tracks and
energy deposits (``clusters") in the electromagnetic and hadron calorimeters.
The tracks and clusters must pass a set of quality requirements
similar to those used in
previous OPAL Higgs boson searches~\cite{pr367}.
Charged particle tracks and energy clusters, satisfying these requirements,
are associated to form ``energy flow objects''.
A matching algorithm is employed to reduce double counting 
of energy in cases where charged tracks point toward 
electromagnetic clusters~\cite{pr367}.
The energy flow objects are then grouped 
into jets and contribute to the total energy and
momentum of the event. The association into jets 
is performed by the Durham jet finder algorithm~\cite{drm}.

For the \ho\Ao\ra hadrons 
analysis, the data are separated into four \sqrts~ bins for the 1999
data, with average \sqrts\ values of approximately 192~GeV, 196~GeV,
200~GeV, and 202~GeV.  All of the 2000 data is treated
together, with an average \sqrts\ of 206~GeV. The luminosities
of each of the five data samples are given in Table~\ref{fihalumitable}.

\begin{figure}
\centerline{\epsfig{file=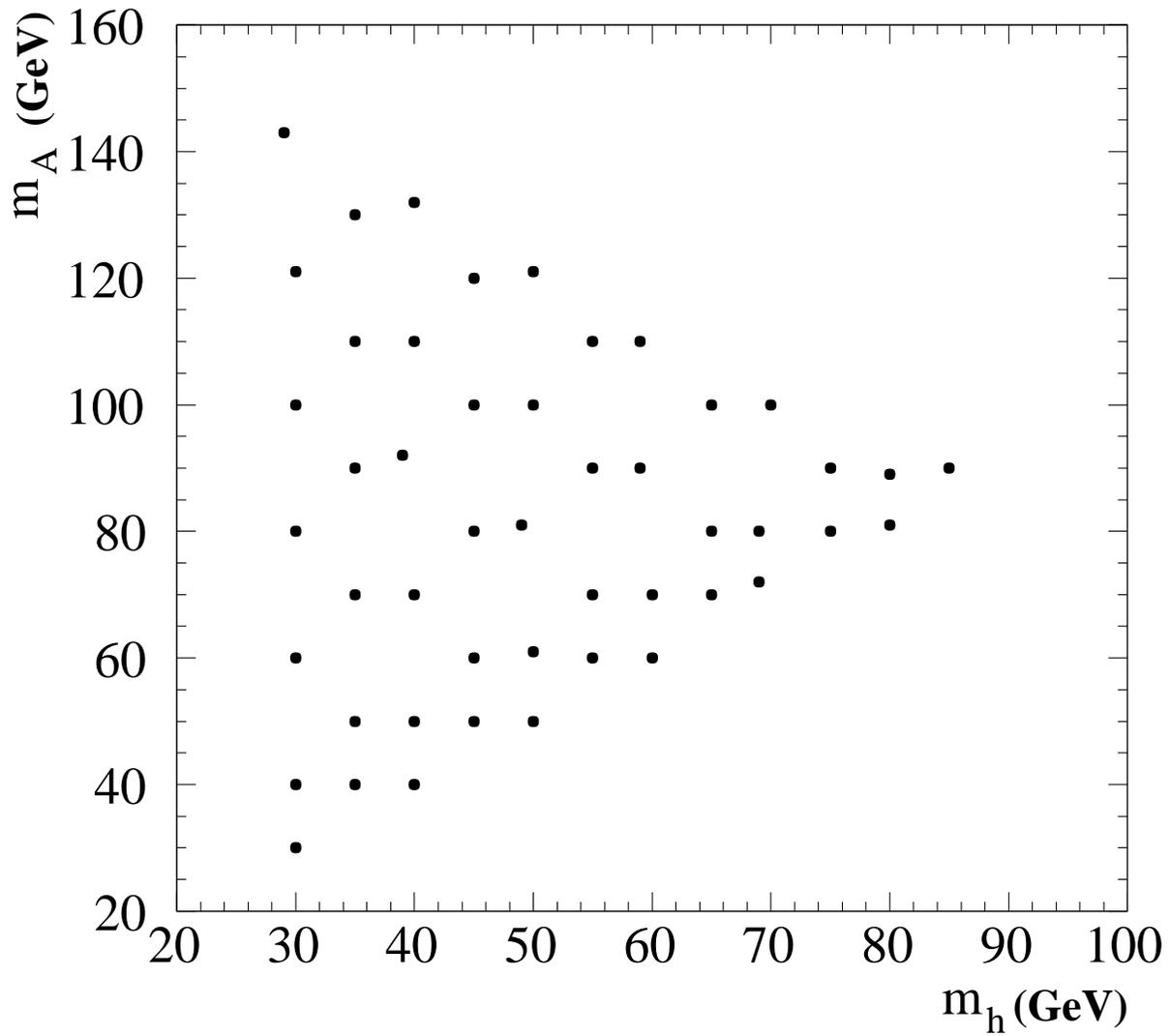,width=\textwidth}}
\caption[]{\label{fihagrid}The Monte Carlo mass grid on which the signal
was generated for the year 2000 analysis. Similar grids are used
for the four 1999 samples.}
\end{figure}

A variety of Monte Carlo samples has been generated in order to estimate the
detection efficiencies for Higgs boson production and background from SM 
processes.
Monte Carlo signal samples are generated using {\tt HZHA}~\cite{hzha}
on a grid in the (\mh, \mA) plane, as shown in Figure~\ref{fihagrid}.
One thousand signal Monte Carlo events are generated and simulated
for each grid point for each of the five centre-of-mass energies 
given in Table ~\ref{fihalumitable}.  
Samples are
generated for four flavour combinations of the decays, \ho\Ao\ra\bb\bb,
\ho\Ao\ra\bb\cc, \ho\Ao\ra\cc\cc, and \ho\Ao\ra~gggg.

For the background processes the following event generators are used: {\tt KK2f}~\cite{kk2f}
for \Zgs\ra\qq($\gamma$), \mm$(\gamma)$ and \tautau$(\gamma)$,
{\tt BHWIDE}~\cite{bhwide} for \ee($\gamma$) and {\tt KORALW}~\cite{koralw} and {\tt grc4f}~\cite{grc4f} for four-fermion
processes. The {\tt KK2f} prediction for \qq($\gamma$) were compared to {\tt PYTHIA}~\cite{pythia}
and {\tt HERWIG}~\cite{herwig} samples. The {\tt KORALW} prediction for hadronic and
semi-leptonic four-fermion processes (with no electron in the final state)
were compared to {\tt grc4f} samples. {\tt JETSET}~\cite{pythia} is used as the principal model
for fragmentation.

The detector response to the generated particles is simulated in full
detail~\cite{gopal}.

\begin{table}[t]
\begin{center}
\begin{tabular}{|l|c|c|}
\hline
Year & \sqrts\ [GeV] & $\mathcal{L}$ [\pb] \\ \hline
1999 & 191.6 & 28.9 \\
1999 & 195.5 & 74.8 \\
1999 & 199.5 & 77.2 \\
1999 & 201.6 & 36.1 \\
2000 & 206.0  & 208 \\ \hline 
\end{tabular}
\caption[]{\label{fihalumitable}
The integrated luminosities, $\mathcal{L}$, at each
centre-of-mass energy, $\sqrt{s}$,  used for the flavour independent 
\ho\Ao\ra~hadrons search.}
\end{center}
\end{table}

\begin{boldmath}
\section{Flavour independent search for \ee\ra\ho\Ao}
\label{sect:fiha}
\end{boldmath}


For some values of the parameters specifying a 2HDM(II), 
\ee\ra\ho\Zo\ is suppressed,
either kinematically or due to small $\sin^2(\beta-\alpha)$.  The decays
\ho\ra\bb\ and \Ao\ra\bb\ may be also suppressed in a subset of these models
because of reduced couplings.  In such models, the largest signal
for Higgs boson production may be \ee\ra\ho\Ao\, where both the \ho\ and 
the \Ao\ decay hadronically, but not necessarily to \bb.  The dominant decay modes
of the Higgs bosons may even be to pairs of gluons.  The final state investigated
here is four well-separated jets of hadrons of any flavour.

The search presented here is based on the search procedure applied to the 
189~GeV data~\cite{2hdmpaper}.  The search is extended by including all data
collected at \sqrts~=~192 to 209~GeV, and by introducing a likelihood
discriminant to combine information carried by several different kinematic
variables which are measured in each event.  The results of the
search for \ee\ra\ho\Ao\ra~hadrons in~\cite{2hdmpaper} are combined with
the results of the newer searches; the older data are not re-analysed.

Without flavour tagging, and without a fixed mass constraint such as the
\Zo\ mass, the assignment of dijets in selected candidate events
to the \ho\ and \Ao\ is ambiguous.
There are six possible assignments of jets to bosons.  The pairing chosen
in this analysis is the same as that used in~\cite{2hdmpaper}.  This
is the pairing which minimizes the $\chi^2$ of a beam energy and
momentum-constrained kinematic fit to the (\mh, \mA) hypothesis under study.
The pairing of each event therefore depends on the test masses.

The Standard Model backgrounds for \ee\ra\ho\Ao\ra~hadrons are large.
One of the main sources is \ee\ra\qq\ (approximately
18~pb for events with less than 20~$\%$ of the centre-of-mass energy
in initial state radiation at \sqrts~=~205~GeV~\cite{cernew}),
and includes events which
may have one or more initial-state-radiation photons.
Much of this background has only two jets.  Hard gluon radiation
in a fraction of \ee\ra\qq\ events produces four-jet final states.
In general, the radiation of a gluon produces jets
close in angle to other jets, and so this background tends to mimic 
\ee\ra\ho\Ao\ra~hadrons where either the \ho\ or the \Ao\ is light.  Conversely
if both the \ho\ and the \Ao\ are assumed to be light, then signal events are
more two-jet-like, and closely resemble a larger fraction
of the \qq\ background.  The
other main source of background is the production of pairs of vector
bosons which decay hadronically.
The most important process is
\ee\ra\WW\ra~hadrons (approximately 8~pb at \sqrts~=~205~GeV~\cite{cernew});
the process \ee\ra\ZZ\ra~hadrons contributes at a 
smaller level (approximately 0.5~pb at \sqrts~=~205~GeV~\cite{cernew}).
These background processes produce four-jet events with
large invariant masses when jets are combined together in pairs.
The \WW\ background mimics a signal with $\mh \approx \mA \approx$ \mWn.  Furthermore, because
the pairing of jets to bosons is ambiguous, the  \WW\ and \ZZ\ backgrounds
can contribute everywhere in the (\mh, \mA) plane.

The event selection starts with a cut-based preselection and
proceeds with a selection based on a likelihood function.\\

\subsection{Preselection}
\label{fihapreselection}

The preselection is based on the preselection used in the
search for the Standard Model Higgs boson in the Higgs-strahlung
process in the four-jet final state~\cite{pr367}, without
the requirement that two jets are consistent with \mZ.  The cuts are:
\begin{enumerate}

\item Events must satisfy the hadronic final-state 
requirement of~\cite{multihadronic}.
The effective centre-of-mass energy, \sprime, obtained
by kinematic fits assuming that initial state radiation photons
are lost in the beampipe or seen in the detector~\cite{multihadronic}
must be at least 80~\% of \sqrts.
The value of the jet resolution parameter, $y_{34}$,
at which an event is reclassified from 3-- to 4--jet event by the Durham algorithm~\cite{drm} must
exceed 0.003.

\item The $C$ parameter, which gives a measure
of the spherical shape of the event~\cite{cparam}, must be larger
than 0.25.

\item The $\chi^2$ probability of a 4-constraint (4C) kinematic fit requiring 
energy and momentum conservation must be greater than $10^{-5}$.

\item The event is forced to have four jets and each of the four jets 
 must have at least one charged particle track.

\item No jet-pairing combination may have a 6-constraint (6C) kinematic fit
probability greater than 0.2, where the fit constrains
the total energy, momentum, and the masses of both dijets to \mW.

\end{enumerate}
Table~\ref{fihacutflow} shows the numbers of events
passing each of the preselection requirements in the data taken
in 1999 and 2000, along with the expected backgrounds and
the efficiency for a signal with $\mh = 50$~GeV and $\mA = 100$~GeV.

\subsection{Likelihood Selection}
\label{fihalikelihood}

The following nine kinematic variables were considered in the
likelihood selection:

\begin{itemize}

\item $\log(\Delta\chi^2(\mh,~\mA))$.  A full description of the
procedure for computing this variable is given in~\cite{2hdmpaper}, where
it was used as the only discriminant variable.  For each event,
a 4C kinematic fit is performed, constraining the event energy and
momentum to the centre-of-mass energy and momentum.  The 
$\Delta\chi^2(\mh,~\mA)$ value is the additional $\chi^2$ incurred
when constraining one pair of jets to have invariant mass \mh\ and
the other to have \mA.  The pairing is chosen to minimise
$\Delta\chi^2(\mh,~ \mA)$.  This is the only variable for which the
value depends on the test mass combination.

\item $|\cos\theta_{\mathrm{thrust}}|$, obtained from the polar angle, 
$\theta_{\mathrm{thrust}}$, of the thrust axis.

\item The event aplanarity $A$.

\item $\log(y_{34})$.

\item The jet-angle sum $J_s$~\cite{alephjas}.  This variable
is the sum of the four smallest dijet angles.

\item $(E_{\mathrm{max}}-E_{\mathrm{min}})/\sqrt{s}$, the difference
between the energy of the highest-energy jet and 
that of the lowest-energy jet, divided
by the centre-of-mass energy.

\item The jet-charge-signed $\cos\theta_{\mathrm W}$.  This variable
is computed using the jet pairing
which maximizes the 6C fit probability to the \WW\ mass hypothesis.
The quantity 
\begin{equation}
Q_j~=~\sum_{i=1}^{n_{\mathrm tracks}}q_i~ 
{\mathrm sign}(p_\parallel^i ) \sqrt{|p_\parallel^i|}
\end{equation}
is computed for each jet $j$, where $n_{\mathrm tracks}$ is the number of
tracks in the jet, $q_i$ is the charge of the $i$th track in the jet, and 
$p_\parallel^i~=~{\vec p}_i\cdot{\hat n}_j$ where $\vec p_i$ is the three-momentum
of the $i$th track in the jet and $\hat n_j$ is the unit vector pointing along the jet axis.
If jets $j$ and $k$ are paired together to form a W boson candidate
and jets $l$ and $m$ are paired to form the other W boson candidate, then 
\begin{equation}
\cos\theta_{\mathrm W}~ = ~\frac{\left({\vec P}_j+{\vec P}_k\right)\cdot {\hat z}}
{\left|{\vec P}_j+{\vec P}_k\right|}{\mathrm sign}\left(Q_j+Q_k-Q_l-Q_m\right),
\end{equation}
where $\vec P_j$ is the three-momentum of the jet $j$ after the
4C-fit and $\hat z$ is the unit vector
pointing along the electron beam axis.

\item $\log(W_{\mathrm CC03})$, the logarithm of the WW matrix element 
calculated by the {\tt EXCA-\\LIBUR} program~\cite{excalibur} using 
the CC03 set of diagrams.  The four-vectors of the jets after the 4C-fit
are used as inputs to the calculation.
The matrix element is computed for all possible assignments of
jet pairs to W bosons and the largest value is used.

\item $\log(W_{\mathrm{QCD}})$, the logarithm of the 
\ee\ra\qq\ra~four-jet matrix element~\cite{cataniseymour}.
The matrix element is computed for all possible permutations of jets
and the largest value is used.

\end{itemize}

Reference histograms are formed for each likelihood input variable,
for each signal grid point, separately for the \ee\ra\qq\ 
background (2f), the \ee\ra~qqqq background (4f), and the expected
signal, accumulating events which pass the preselection requirements.
The \ellell\qq\ and $\ell\nu_\ell$qq backgrounds are expected to be 
small after the preselection -- their numbers are included in the
4f background numbers in Table~\ref{fihacutflow}.
They do not contribute to the reference histograms, but are
accounted for in the final background estimates in the likelihood output
histograms.  The backgrounds from two-photon processes are negligible after
the preselection. 

\begin{figure}
\begin{center}
{\Large\bf{OPAL}}
\end{center}
\centerline{\epsfig{file=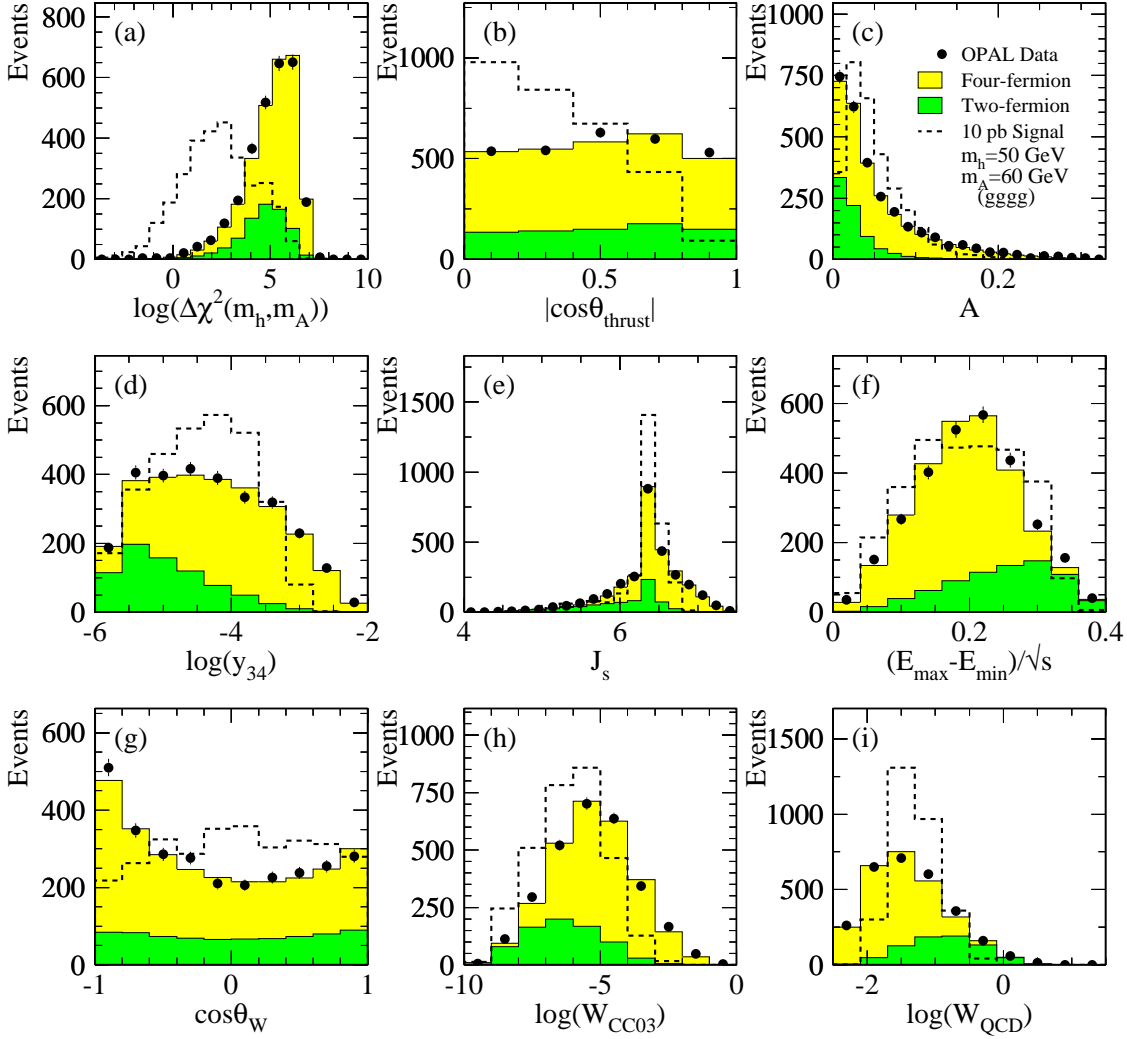,width=\textwidth}}
\caption[]{\label{fiharef}  
 Distributions of the likelihood input
variables for preselected events in the flavour-independent hadronic
\ho\Ao\ search:  
(a) $\log(\Delta\chi^2(\mh,\mA))$, 
(b) $|\cos\theta_{\mathrm{thrust}}|$, 
(c) the event aplanarity $A$,
(d) $\log(y_{34})$,
(e) $J_s$, 
(f) $(E_{\mathrm{max}}-E_{\mathrm{min}})/\sqrt{s}$, 
(g) the jet-charge-signed $\cos\theta_{\mathrm W}$, 
(h) log(W$_{\mathrm{CC03}}$), and
(i) log(W$_{\mathrm{QCD}}$).
The dark-shaded histograms show the two-fermion (\qq ) background, the
light-shaded histograms show the four-fermion background.  The backgrounds
are shown stacked atop one another. The dashed
histograms show a 10~pb signal of \ee\ra\ho\Ao\ra~gggg with 
$\mh=50$ GeV and $\mA=60$ GeV, and the
points with error bars show the OPAL data collected in 1999 and 2000.
}
\end{figure}

Figure~\ref{fiharef} shows the distributions of the likelihood
input variables for the data collected in 1999 and 2000, the corresponding
background estimate, and the expected signal for a fully gluonic decay
for the hypothesis \mh~=~50~GeV, \mA~=~60~GeV.

For any point in the (\mh, \mA) plane within kinematic reach of the
LEP beam energy, and with $\mh>30$~GeV and $\mA>30$~GeV, a separate
likelihood function may be constructed from the reference histograms
of the input variables.  These are formed by interpolating
the signal reference histograms using nearby Monte Carlo
grid points. The background reference histograms
must be interpolated also for the $\log(\Delta\chi^2(\mh, ~\mA))$ variable.
These interpolations make use
of the method described in~\cite{pvmorph}, extended to interpolate
histograms which are functions of two variables, \mh\ and \mA.
The likelihood output histograms are also
interpolated, separately for the signal and each background
contribution, but each bin's contents is linearly interpolated
between the Monte Carlo grid points.
The interpolated reference histograms allow the
computation of the value of the likelihood function for each
candidate for each point in the (\mh, \mA) plane, and the
interpolated signal and background likelihood histograms
allow comparison of the data likelihood distribution with the
signal and background predictions.

Signal events with \ho\Ao\ra\bb\bb\ are easier to separate
from the background than signal events with \ho\Ao\ra~gggg
since the gggg case has a poorer reconstructed mass resolution
which deteriorates the $\log(\Delta\chi^2(\mh,~\mA))$ likelihood variable.
The signal reference histograms are created using 
samples of \ho\Ao\ra\bb\bb\ signal Monte Carlo events, and the
signal likelihood histograms are filled with \ho\Ao\ra~gggg
events, ensuring the statistical independence of the
reference and likelihood histograms, and also ensuring the
conservativeness of the performance over the possible final
states of the \ho\ and \Ao\ decays.

\begin{table}[t]
\begin{center}
\begin{tabular}{|r|r|r|r|r|r|}
\hline
Cut & 2f bkg & 4f bkg & tot bkg & data & eff~[\%]\\\hline
 (1) &  1031.6&  3206.4&  4238.0&  4479& 76.6\\
 (2) &   976.0&  3203.1&  4179.1&  4436& 76.6\\
 (3) &   887.3&  2977.2&  3864.5&  3807& 71.0\\
 (4) &   784.7&  2820.2&  3605.0&  3580& 70.4\\
 (5) &   754.3&  2035.7&  2789.9&  2835& 68.5\\\hline
\end{tabular}
\caption[]{\label{fihacutflow}  
The numbers of events
passing each of the preselection requirements in the data taken
in 1999 and 2000, along with the expected backgrounds and
the luminosity-weighted average efficiency for a signal with $\mh~=~50$~GeV, $\mA~=~100$~GeV.}
\end{center}
\end{table}

\begin{figure}
\centerline{\epsfig{file=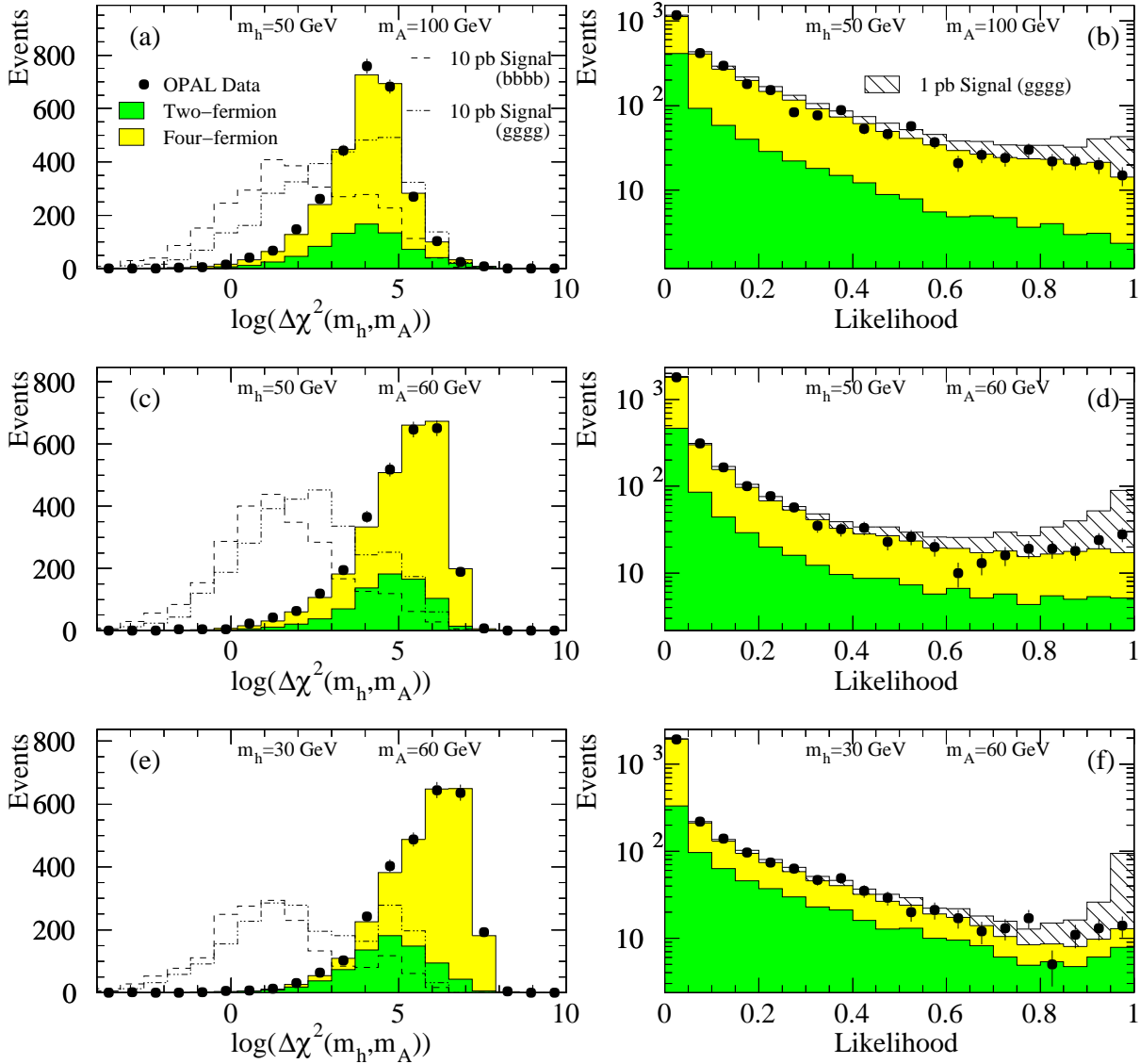,width=\textwidth}}
\caption[]{\label{fihalike}Distributions of the $\log(\Delta\chi^2(\mh,\mA))$
variable and the likelihood output variable for three test-mass hypotheses,
(\mh, \mA)=(50~GeV,~100~GeV), (50~GeV,~60~GeV), and (30~GeV,~60~GeV).
OPAL data are shown with the points, light-shaded histograms show the
four-fermion Standard Model background expectations, and dark-shaded histograms
show the two-fermion (\qq ) Standard Model background expectations.  The two background
rates are shown stacked in all histograms.  In the likelihood
histograms, a 1~pb signal assumed to decay to gggg is shown with the hatched
histograms, added on top of the background sum.  For the $\log(\Delta\chi^2(\mh,\mA))$
distributions, a 10~pb signal assumed to decay to \bb\bb\ is shown with dot-dashed
histograms which are not added to the background, while a 10~pb signal 
assumed to decay to gggg is shown with dotted histograms, also not added to the
background.
}
\end{figure}

Example likelihood distributions are shown in Figure~\ref{fihalike}
for all data collected in 1999 and 2000, along with the SM background
expectations and signal expectations,
for three test-mass hypotheses, (\mh, \mA)~=~(50~GeV,~100~GeV),
(50~GeV,~60~GeV), and (30~GeV,~60~GeV).  The corresponding
distributions of $\log(\Delta\chi^2(\mh,~ \mA))$ are also shown
to illustrate how the distributions of the 
signal, the expected backgrounds, and the candidates
change with the test-mass hypothesis.

The distribution of the likelihood is used directly as the input to the limit
calculation.  In the presence of systematic uncertainties
on the background rate, including bins with low expected
signal-to-background ratios reduces the sensitivity of the
search.  A lower cut on the likelihood variable of 0.8 is chosen,
independent of the test mass hypothesis, in order to improve
the sensitivity of the search.  Table~\ref{fihaliketab} lists
the numbers of events passing the likelihood cut for each
of the test masses on the Monte Carlo grid, the expected
backgrounds, and the expected signal efficiencies for the \ho\Ao\ra~gggg
decay hypothesis.  The signal efficiencies are also calculated separately
for the \cc\cc, \bb\bb, and the \bb\cc\ decay hypotheses.  For nearly all
test mass combinations, the \ho\Ao\ra~gggg hypothesis yields the least
efficiency, and for the remainder, the differences are within the
uncertainties.

\begin{table}[H]
\begin{center}
\vspace{0.5cm}
\begin{small}
\begin{tabular}{|r|r|r|r|r|r|r|r|r|r|}
\hline
\mh\  & \mA\  & 2f  & 4f  & total & data & eff $[\%]$ & eff $[\%]$ & eff $[\%]$ & eff $[\%]$ \\
$[\G]$ & $[\G]$ & bkg & bkg &  bkg  & $~$     & gggg     & \cc\cc  & \bb\bb   & \bb\cc \\
\hline
  30.0& 30.0& 27.0&  6.1& 33.1$\pm$3.1& 20&  1.9$\pm$0.5 &   2.1 & 1.7  & 1.0\\
  40.0& 30.0& 42.1& 11.8& 53.9$\pm$5.1& 42& 12.1$\pm$1.4 &  13.7 & 12.2 &13.9\\
  60.0& 30.0& 24.1& 15.2& 39.2$\pm$3.7& 43& 26.6$\pm$2.4 &  27.7 & 30.7 &29.6\\
  80.0& 30.0& 16.6& 22.7& 39.3$\pm$3.7& 46& 22.0$\pm$2.1 &  30.1 & 27.4 &32.4\\
 100.0& 30.0& 15.3& 48.2& 63.5$\pm$6.0& 69& 16.4$\pm$1.7 &  22.9 & 24.1 &22.9\\
 120.0& 30.0& 12.7& 34.6& 47.3$\pm$4.4& 59&  8.8$\pm$1.1 &  20.1 & 17.9 &24.3\\
 140.0& 30.0& 11.3& 19.3& 30.6$\pm$2.9& 37&  3.7$\pm$0.7 &   8.3 &  8.8 & 7.7\\
  40.0& 40.0& 29.5& 15.1& 44.6$\pm$4.2& 32& 22.1$\pm$2.1 &  22.7 & 22.3 &19.1\\
  50.0& 40.0& 33.0& 32.1& 65.1$\pm$6.1& 64& 36.1$\pm$3.0 &  36.4 & 33.6 &36.0\\
  70.0& 40.0& 19.4& 34.5& 53.8$\pm$5.1& 46& 26.7$\pm$2.4 &  35.8 & 34.6 &33.7\\
  92.0& 40.0& 15.7& 59.1& 74.8$\pm$7.0& 82& 17.8$\pm$1.8 &  28.4 & 24.7 &26.4\\
 110.0& 40.0& 14.9& 58.2& 73.1$\pm$6.9& 74& 14.7$\pm$1.6 &  22.9 & 25.7 &22.6\\
 130.0& 40.0& 18.7& 47.6& 66.4$\pm$6.2& 79&  8.7$\pm$1.1 &  12.6 & 14.5 &15.7\\
  50.0& 50.0& 18.1& 32.3& 50.3$\pm$4.7& 54& 36.5$\pm$3.1 &  39.6 & 41.6 &41.2\\
  60.0& 50.0& 20.9& 49.5& 70.4$\pm$6.6& 89& 34.0$\pm$2.9 &  37.9 & 36.5 &33.6\\
  80.0& 50.0& 11.9& 57.2& 69.1$\pm$6.5& 58& 23.6$\pm$2.2 &  31.7 & 32.2 &30.5\\
 100.0& 50.0& 12.6& 66.7& 79.3$\pm$7.5& 79& 16.7$\pm$1.7 &  25.2 & 25.5 &26.0\\
 120.0& 50.0& 18.5& 59.9& 78.4$\pm$7.4& 78& 13.8$\pm$1.5 &  19.6 & 20.4 &22.7\\
  60.0& 60.0& 15.2& 43.0& 58.2$\pm$5.5& 61& 31.1$\pm$2.7 &  37.5 & 40.0 &39.7\\
  70.0& 60.0& 16.8& 64.0& 80.8$\pm$7.6& 87& 26.3$\pm$2.4 &  33.5 & 33.2 &37.3\\
  90.0& 60.0& 10.5& 75.6& 86.2$\pm$8.1& 88& 17.4$\pm$1.7 &  23.5 & 23.6 &23.6\\
 110.0& 60.0& 20.1& 79.4& 99.5$\pm$9.4& 92& 16.3$\pm$1.7 &  22.3 & 20.9 &21.5\\
  70.0& 70.0&  8.4& 50.7& 59.1$\pm$5.6& 57& 22.5$\pm$2.1 &  36.0 & 31.5 &34.6\\
  80.0& 70.0&  8.7& 99.2&107.9$\pm$10.1&106& 18.1$\pm$1.8 &  27.2 & 29.8 &28.0\\
 100.0& 70.0& 12.0& 84.1& 96.1$\pm$9.0&100& 16.4$\pm$1.7 &  23.3 & 24.9 &21.2\\
  80.0& 80.0&  1.7& 56.9& 58.6$\pm$5.5& 54&  2.5$\pm$0.5 &   4.6 &  9.6 & 6.1\\
  90.0& 80.0&  4.4& 92.9& 97.2$\pm$9.1&110& 11.5$\pm$1.3 &  18.2 & 19.5 &14.6\\
  90.0& 90.0& 21.0&142.0&162.9$\pm$15.3&162& 25.7$\pm$2.3 &  31.7 & 32.5 &29.8\\
\hline
\end{tabular}
\end{small}
\caption[]{\label{fihaliketab}
The expected Standard Model backgrounds, observed data counts, and
expected signal efficiencies for the flavour independent \ho\Ao\ search
as a function of the test mass hypotheses.  All data collected in 1999
and 2000 are combined, and the signal efficiencies are luminosity-weighted
averages.  The efficiencies are listed separately for gggg, \cc\cc, \bb\bb\
and \bb\cc\ decay hypotheses. The errors given in table are statistical 
and systematic added in quadrature.}
\end{center}
\end{table}

\subsection{Systematic Uncertainties}

  Because of the cut on the likelihood of 0.8, the systematic
uncertainties are evaluated only for the numbers of events
passing this likelihood cut.   Correlations between the uncertainties on
the signal and background rates, as well as between samples
taken at different centre-of-mass energies are evaluated and used in the computation
of the confidence levels.

\begin{itemize}

\item {\bf Monte Carlo Statistics}  For a typical point in the Monte
Carlo test mass grid, the MC statistical error on the signal rate is 7~\%,
but it is larger for models for which both \mh\ and \mA\ are small, due
to the lower signal efficiency for such models.
For nearly all model points, the MC statistical
uncertainty on the background rate is between 1.5~\% and 3~\%, but
for signals when either \mh\ or \mA\ is low, it can be as large as 5~\%.
An overall uncertainty of 7~\% and 3~\% is taken
for the signal and background rates, respectively.

\item {\bf Jet Energy Resolution}  The jet energy resolution is
uncertain in the barrel region by about 3--5~\%, but this uncertainty
is approximately 15~\% in the endcap regions.  Uncertain jet energy
resolution results in approximately a 5~\% uncertainty on both the
signal and background rates.

\item {\bf Jet Energy Scale}  The jet energy scale is uncertain at
the 1~\% level.  The corresponding uncertainty
on the rate of events passing the likelihood cuts is approximately 2~\%
for both the signal and the background.

\item {\bf Jet Angle Resolution}  A 22~mrad angle resolution
uncertainty in both $\theta$ and $\phi$ results in an uncertainty
on the rate of events passing the likelihood cuts that is approximately 1~\%
for both the signal and the background.

\item {\bf Interpolation Uncertainty}
To test the reliability of the interpolation procedure, a single Monte Carlo
signal point on the grid is deleted, and the interpolation procedure is
used to replace it, and the process is repeated for all MC grid points
not on the edges.  Using the differences found in the selection rates,
an uncertainty of 3.5~\% is assigned to the signal efficiency
and an uncertainty of 4~\% is assigned to the background rate
due to interpolation errors.

\item {\bf Four-Fermion Cross-Section}  Nearly all of the background
at model points near the expected limit is from four-fermion production.
It is dominated by \ee\ra\WW\ production, although \ee\ra\ZZ\ contributes as
well.  A 2~\% uncertainty is assessed on the production cross-section
of these events~\cite{4ferror}.

\item {\bf Monte Carlo Background Sample Comparison}  The {\tt KK2F}
Monte Carlo generator using {\tt PYTHIA} as the fragmentation and hadronisation
model was used to generate the central values
for the \qq\ background rates.  These rates were compared
to the same {\tt KK2F} sample but re-hadronised with {\tt HERWIG}, and also
with a sample generated entirely with {\tt PYTHIA}.
{\tt KORALW} was used to generate the qqqq, the \qq\ellell\ and
the qq$\ell\nu_\ell$ background rate
central values, and {\tt grc4f} was used as the comparison generator.
An uncertainty of 5.4~\% is assessed on the background rates
passing the likelihood cuts.

\end{itemize}

  When all uncertainties are added in quadrature, the systematic
error on the background rate is 9.4~\% and the systematic error on
the signal rate is typically 9.6~\%, with larger values for the signal systematic
uncertainty for models with low efficiency, due to the Monte Carlo statistical
uncertainty. 
The correlations between the signal and
background uncertainties and between years are 
taken into account. The Monte Carlo statistical error affects the signal and background predictions
and are uncorrelated between energies, and signal and background.
The other uncertainties are correlated between centre-of-mass energies.

\section{Confidence level calculation}
\label{sect:combmet}

Following the statistical method described
in \cite{tomnim, readst}, the direct searches listed in Section~\ref{sect:datasamples}
are combined to increase 
the Higgs boson discovery potential and, in case of absence of 
signal, the exclusion power.

The confidence levels are derived from a test statistic, $Q$, which is
defined such that $Q$ quantifies the compatibility of the data with two hypotheses:
a) the background hypothesis, and b) the signal+background hypothesis.
The confidence levels are computed from a comparison of the observed test statistic and 
its probability distributions for a large number of simulated experiments for these two hypotheses.
The results of the different search channels are expressed 
in bins of discriminating variables defined in the individual searches ({\it{e.g.}} mass, likelihood,
neural network output, etc.).
The ratio $Q={\cal L}_{{\mathrm{s+b}}}/{\cal L}_{{\mathrm{b}}}$ of the binned likelihoods for the 
two hypotheses is chosen as the test statistic.
The confidence level for the background hypothesis,
${\mathrm{CL}}_{\mathrm{b}}$,
is defined as the probability to obtain
values of $Q$ no larger than the observed value $Q_{\mathrm{obs}}$, given a large number of hypothetical
experiments with background processes only, 
$\mathrm{CL}_{\mathrm{b}}=P(Q~\le~ Q_{\mathrm{obs}}|{\mathrm{background}}).$
Similarly, the confidence level for the signal+background hypothesis, ${\mathrm{CL}}_{{\mathrm{s+b}}}$,
is defined as the probability to obtain
values of $Q$ not larger than observed, given a large number of hypothetical
experiments with signal and background processes,
${\mathrm{CL}}_{{\mathrm{s+b}}}=P(Q~\le ~Q_{\mathrm{obs}}|{\mathrm{signal+background}}).$
In principle, ${\mathrm{CL}}_{{\mathrm{s+b}}}$ could be used to exclude the signal+background 
hypothesis, given a model for Higgs boson production. 
However, this procedure may lead to cases when
a downward fluctuation of the background would allow hypotheses to be excluded
for which the experiment has no sensitivity due to the small expected signal rate. Therefore
the ratio 
${\mathrm{CL}}_{\mathrm{s}}={\mathrm{CL}}_{\mathrm{s+b}}/{\mathrm{CL}}_{\mathrm{b}}$ is used. 
It is always greater than ${\mathrm{CL}}_{{\mathrm{s+b}}}$ and the limit obtained in this way is 
thus conservative. We adopt this quantity for setting exclusion limits and consider a hypothesis to
be excluded at the $95\,\%$ confidence level if the corresponding value of ${\mathrm{CL}}_{\mathrm{s}}$ is less than 0.05.

The expected confidence levels are obtained by replacing the observed data with
a large number of simulated events for the background only or signal+background hypotheses.

The effect of systematic uncertainties for the individual channels is 
calculated using a Monte Carlo technique. 
The signal and background estimations are varied within the bounds of the systematic uncertainties, assuming
Gaussian distributions of the uncertainties. Correlations are taken into account.
These variations
are convoluted with the Poisson statistical variations of the assumed signal and background rates
in the confidence level calculation.
The effect of systematic uncertainties on the exclusion limits generally turns out to be small.



\section{Model independent interpretation}
\label{sect:modind}
\begin{figure}
\centerline{\epsfig{file=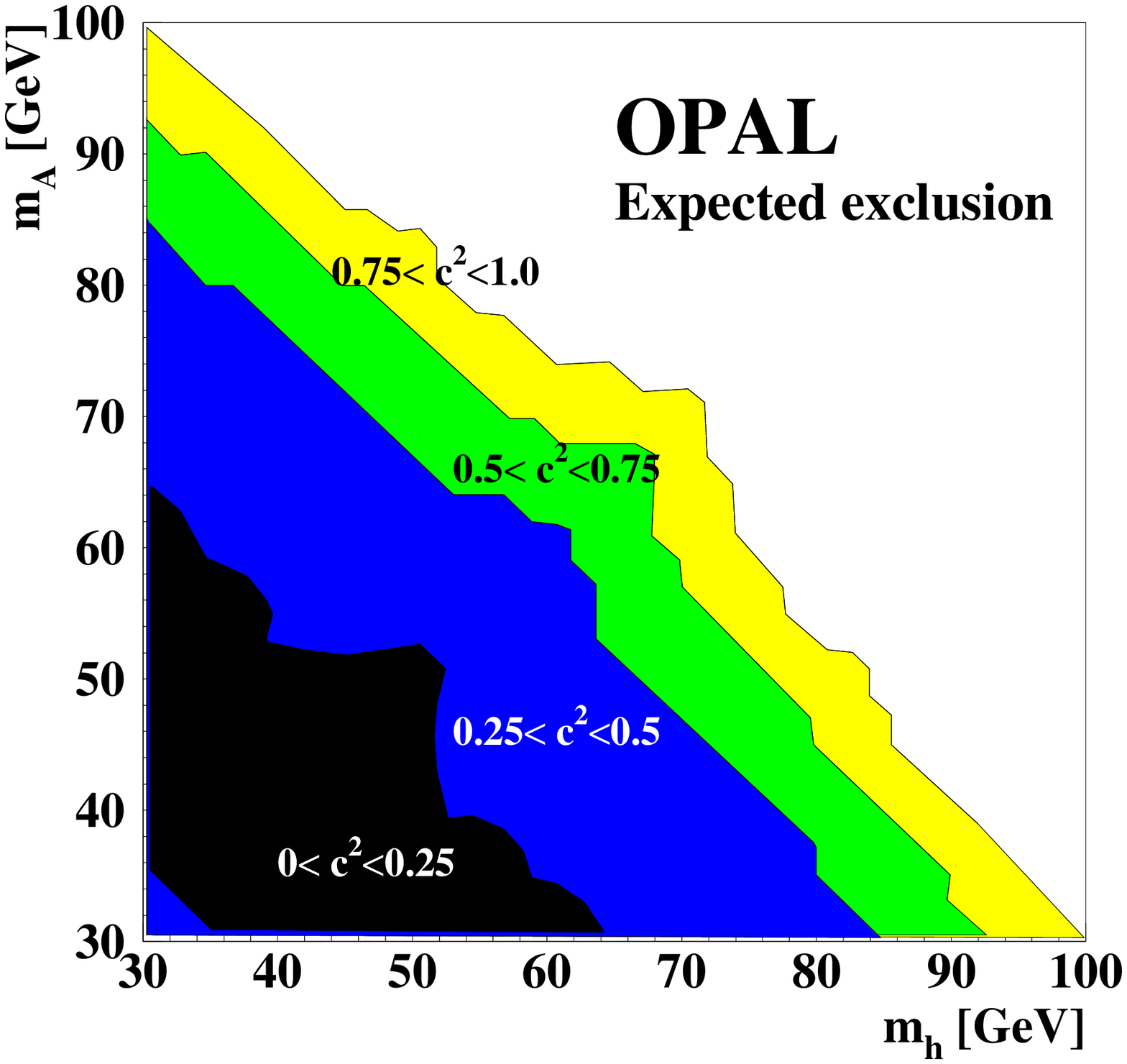,width=\textwidth}}
\caption[]{\label{medex} 
The median expected limit on the coupling $c^2$.
Both the \ho\ and the \Ao\ are assumed to
decay hadronically 100\% of the time.   
}
\end{figure}

\begin{figure}
\centerline{\epsfig{file=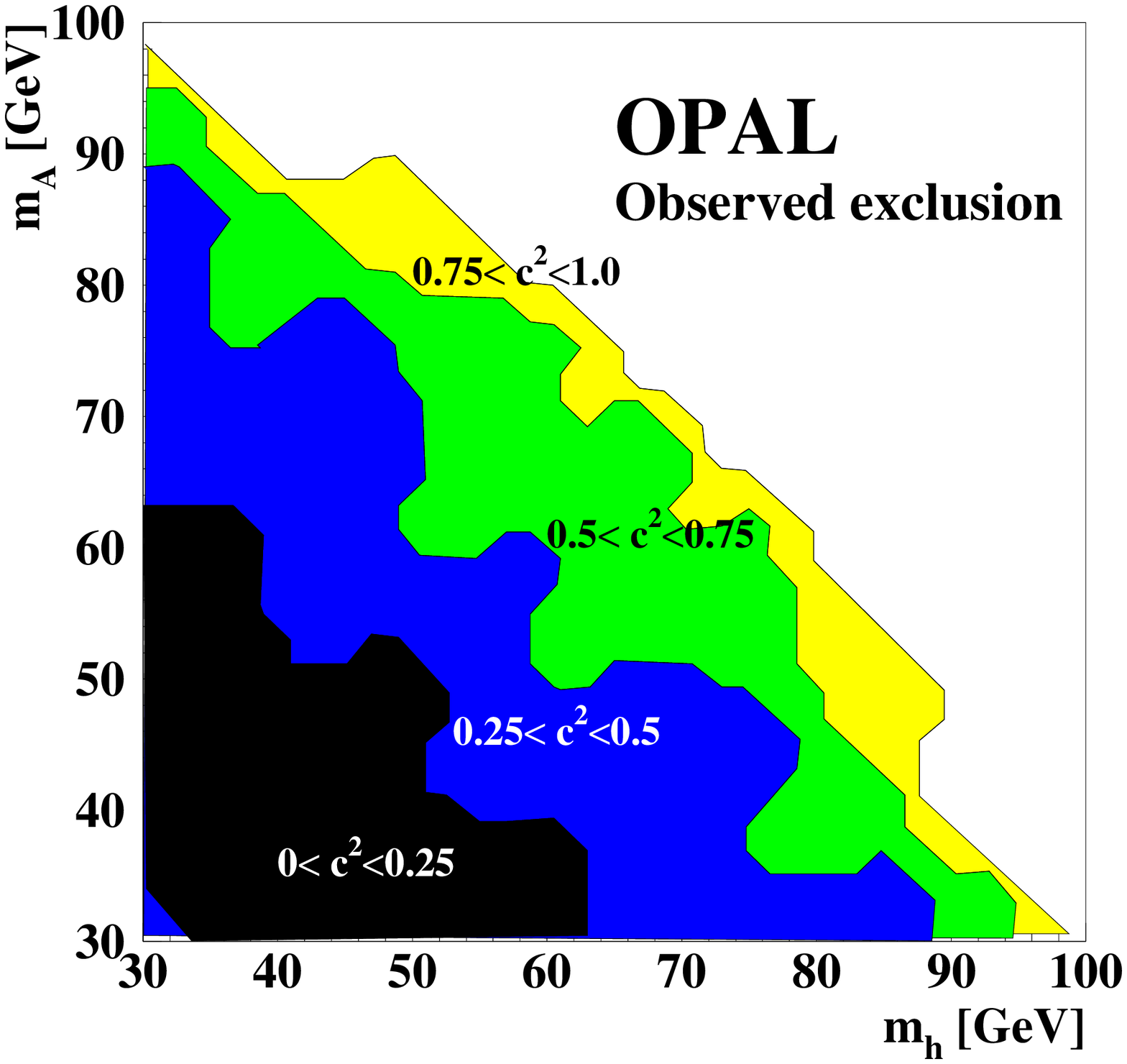,width=\textwidth}}
\caption[]{\label{fihas95}  
The limit on the coupling $c^2$.
Both the \ho\ and the \Ao\ are assumed to
decay hadronically 100\% of the time.}
\end{figure}

Since no excess of data has been observed,
the flavour independent search for \ee\ra\ho\Ao, 
described in Section~\ref{sect:fiha}, is used 
to set 95 $\%$ CL upper limits on the \ho\Ao\ production 
cross-section assuming 100 $\%$ hadronic branching ratios.  
The cross--section for \ee\ra\ho\Ao\ is determined by \mh, \mA,
and the scale factor $c^2$, analogous to the $\cos^2(\beta-\alpha)$
factor of 2HDMs. The scale factor $c^2$ is defined as the ratio
of the pair--production cross--section in the model considered 
and the pair--production cross--section given in Eq. (2) with 
$\cos^2(\beta-\alpha)=1$.
The coupling limit is calculated by finding
the value of $c^2$ for which $\cls = 0.05$, assuming both 
the \h\ and the \Ao\ to decay 100 \% hadronically.
Figures~\ref{medex} and ~\ref{fihas95} show the 95\%
median expected and observed upper limit on $c^2$, respectively, 
as a function of the test-mass hypotheses, 
(\mh, \mA).

The search for \ee\ra\ho\Ao\ra hadrons is sensitive
in the region where the production cross-section exceeds 200 fb
for \sqrts\ between 189~GeV and 206~GeV.
For 100\% decays to hadrons and $c^2 = 1$, this corresponds to
$\mh+\mA\sim 130$~GeV; for these test masses, the production 
cross-section does
not change significantly between the centre-of-mass energies of the
data used.  Monte Carlo signal samples were generated with
$\mh+\mA$ up to 170 GeV, but due to the low signal cross-sections
for high mass Higgs bosons, the $1-\clb$ results are reported only
up to $\mh+\mA = 145$~GeV.  The higher-mass Monte Carlo samples are
needed to interpolate the reference and likelihood histograms over
the entire range over which the results are produced.

  The composition of the background which passes
the likelihood cut depends strongly on the test-masses \mh\ and \mA.
For low \mh\ and \mA, the 2f background dominates, but for models
near the limit, the 4f background is the most important.  The signal
efficiency after the likelihood cut also depends strongly on the
mass hypotheses.  For $\mh\sim\mA\sim 30$~GeV, the 2f background
is quite large and closely mimics the signal.  Only a very small
fraction of the signal and of the background passes the likelihood
selection requirement because of the reduced separation power
between the signal and the background -- the signal efficiency for
this particular model is only $1.9\pm 0.5\%$.  This small efficiency
is compensated by the large expected signal cross section for
low Higgs masses.  Another mass hypothesis where the signal efficiency
is low is $\mh\sim\mA\sim 80$~GeV.  In this case, the \WW\ background
is dominant, and the separation between the signal and the background
is poor.  For this model, the signal cross-section is between 70 fb
at \sqrts~=~189~GeV and 100~fb at \sqrts~=~206~GeV and is therefore
beyond the range of sensitivity of the search.  Over much of the range
of tested mass hypotheses,  the signal efficiency is between
20\% and 30\%, and signal cross-sections as low as 200~fb are excluded.

\begin{figure}
\centerline{\epsfig{file=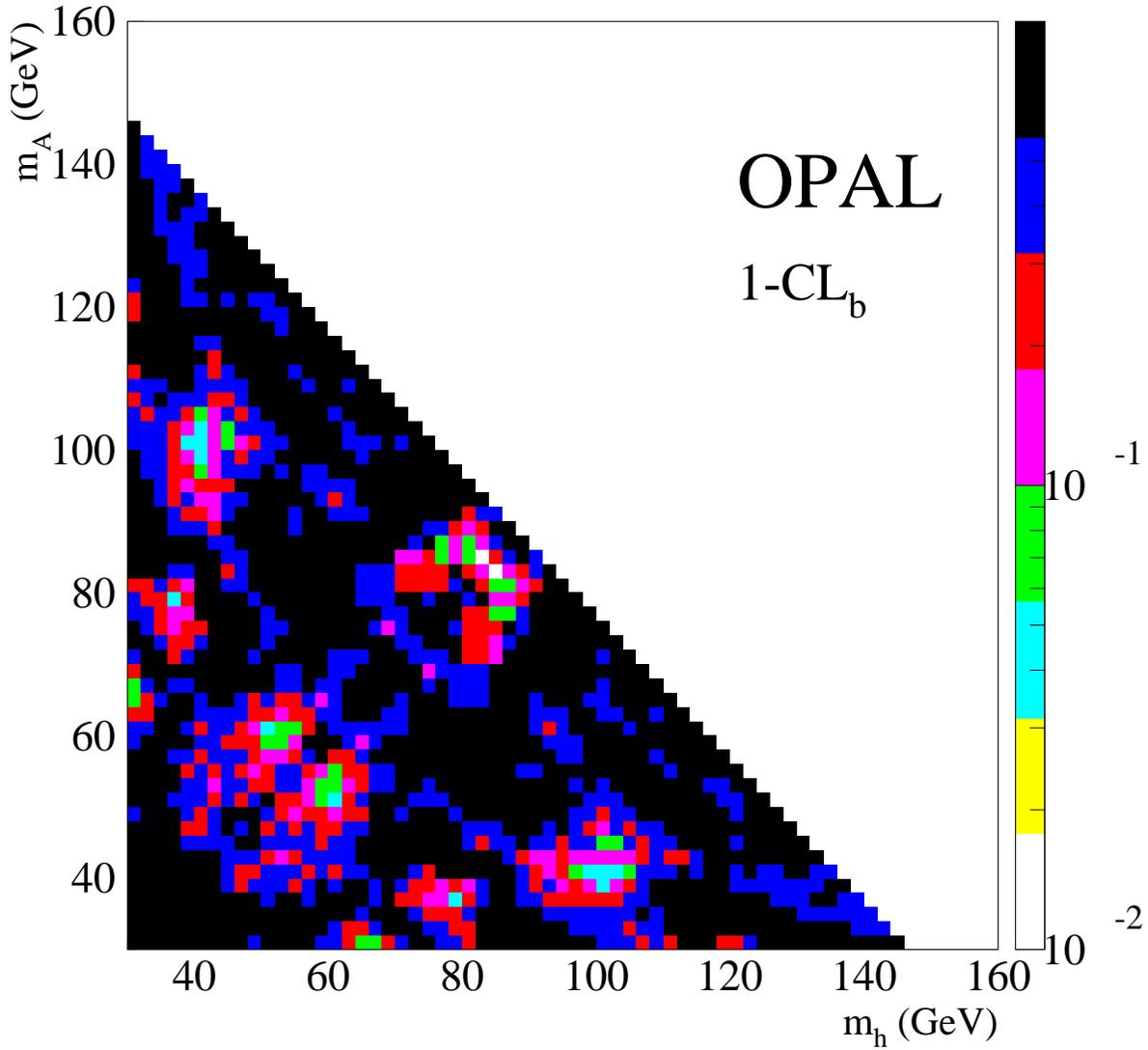,width=\textwidth}}
\caption[]{\label{fihaclbm1}  The value of $1-\clb$ as a function 
of (\mh, \mA). No model has $1-\clb < 0.01$, and
one or more excesses with significance  $1-\clb < 0.01$ are expected
due to the large number of independent searches conducted.}
\end{figure}

To determine whether a signal was observed, we compute $1-\clb$
for each point in the (\mh, \mA) plane in the search region.
Given the mass resolution
of approximately 3~GeV for the sum of \mh\ and \mA, and approximately
7~GeV for the difference between \mh\ and \mA, there are approximately 160
independent searches each of which may have an excess or deficit,
diluting the significance of $1-\clb$.
Thus, more than one independent excess in $1-\clb$ is expected
at the percent level. Figure~\ref{fihaclbm1} shows
$1-\clb$ on a logarithmic scale as a function of \mh\ and \mA.
Nowhere is $1-\clb$ below 1\%.

As a result of the improved sensitivity of the analysis
and the inclusion of data taken at higher centre-of-mass energies
in the years 1999 and 2000, the excluded domains in Figure~\ref{fihas95} are 
extended substantially beyond those obtained in \cite{2hdmpaper}.

\section{Search channels and external constraints used in the 2HDM(II) interpretation}
\label{sect:datasamplesall}

\subsection{Data samples and final state topologies studied }
\label{sect:datasamples}
\label{subsect:CL}
The present study relies on the data collected by 
OPAL at  $\sqrts \approx \mZ$ and from $\sqrts \approx 183$ GeV to 209 GeV, the highest 
\ee\ collision energy attained at LEP. 
This paper uses existing published analyses for all
but the new \h\A\ flavour independent channel described in Section~\ref{sect:fiha}.
Channels that use b-tagging
provide useful information in regions of the 2HDM(II) parameter space where 
the Higgs bosons are expected to decay predominantly into \bb\ pairs.
Flavour independent channels, not using any b-tagging information,
are also included in the combination in order to explore
the regions at low $\alpha$~or low \tanb, where the decays of the \h\ and \A\ bosons 
into \bb\ and \tautau pairs are suppressed.
In Table ~\ref{dataused} the 
references to the published OPAL papers 
for the direct search channels combined in the present 2HDM(II) interpretation
are given, together with the corresponding centre-of-mass energies at which
they were performed.

The channels used at $\sqrts \approx \mZ$, 183 and 189 GeV 
are the same as in \cite{2hdmpaper}. 
The integrated luminosities, the numbers of candidate events, the expected 
SM backgrounds and the efficiencies for each
of the b-tagging (flavour independent) \ho\Z\ channels at $192 \le \sqrts \le 209$ GeV
are given in Table~\ref{tab:smflow} (Table~\ref{tab:flavflow}).

The detection
efficiencies quoted in Tables~\ref{tab:smflow} and ~\ref{tab:flavflow}
are given as examples for specific values of \mh.

The integrated luminosities, the numbers of candidate events, the expected 
SM backgrounds and the efficiencies for the most relevant
b-tagged \ho\Ao\ channels are given in Table ~\ref{tab:ahflow}.  

\begin{table}[h]
\begin{center}
\begin{small}
\renewcommand{\arraystretch}{1.2}
\begin{tabular}{|l|c|c|c|c|} 
\hline
& \multicolumn{4}{|c|}{\bf\boldmath $\sqrt{s}$ [GeV]} \\
\hline
Decay topologies & \mZ & 183  & 189  & 192 \ra\ 209  \\
\hline
\hline
\multicolumn{5}{|c|}{\bf\boldmath \h\Z} \\
\hline
\hline
\bb\qq, \bb\nn, \bb\ee, \bb\mm, \bb\tautau            & NA   & \cite{2hdmpaper} & \cite{2hdmpaper} & \cite{pr367}\\
\hline
\qq\nn, \qq\ee, \qq\mm, \qq\tautau                    & \cite{2hdmpaper} & NA   & \cite{2hdmpaper} & \cite{flavind081}\\
\hline
\tautau\qq                                            & \cite{2hdmpaper} & \cite{2hdmpaper}& \cite{2hdmpaper} & \cite{flavind081} \\
\hline
\qq\qq                                      & NA         & NA   & \cite{2hdmpaper} & \cite{flavind081}\\
\hline
\hline
\multicolumn{5}{|c|}{\bf\boldmath \ho\Ao} \\
\hline
\hline
\qq\tautau or \tautau\qq                              & \cite{2hdmpaper} &  NA & NA   & NA  \\
\hline
\bb\bb, \bb\tautau or \tautau\bb                      &  NA  & \cite{2hdmpaper} & \cite{2hdmpaper} & \cite{mssm020}\\
\hline
\qq\qq                                                &  NA  & NA  & \cite{2hdmpaper} &  This paper \\
\hline
\hline
\multicolumn{5}{|c|}{\bf\boldmath \mh\ $\ge$ 2\mA, \ho\Ao\ra\Ao\Ao\Ao} \\
\hline
\hline
\bb\bb\bb                                             &  \cite{2hdmpaper}  & \cite{2hdmpaper} & \cite{2hdmpaper} & \cite{mssm020} \\
\hline
\hline
\multicolumn{5}{|c|}{\bf\boldmath \mh$\ge$ 2\mA\ and \mA$~\le~$2\mb, \h\Z\ra\A\A\Z } \\
\hline
\hline
\A\ra\cc, \tautau, \glgl\ and \Z\ra\nn, \mm, \ee      & NA     & NA   & \cite{lowma058}& \cite{lowma058} \\
\hline
\hline
\multicolumn{5}{|c|}{\bf\boldmath \mh\ $\ge$ 2\mA, \ho\Zo\ra\Ao\Ao\Zo} \\
\hline
\hline
\bb\bb\qq, \bb\bb\nn                                  &  NA & \cite{2hdmpaper} & \cite{2hdmpaper} &  \cite{mssm020}\\
\hline
\bb\bb\ee, \bb\bb\mm, \bb\bb\tautau                   & NA  & \cite{2hdmpaper} & \cite{2hdmpaper} & NA \\ 
\hline
\end{tabular}
\end{small}
\caption[]{\label{dataused}
Direct search channels combined in the present interpretation
of the 2HDM(II). The searches with \qq\ final states include  
\glgl\ production as well. The numbers in the table give the references
to the OPAL publications where a full description of the channel can be found.
Channels marked NA do not exist.
The \h\A\ra\qq\qq\ (\glgl\glgl, \glgl\qq) analysis 
for the data taken in the years 1999 and 2000 is new 
and is described in Section~\ref{sect:fiha} of this paper.}
\vspace{0.5cm}
\end{center}
\end{table}

When scanning the parameter space the efficiency 
is calculated for each point in the (\mh, \mA)
plane for each of the final states considered.

\begin{table}
\vspace*{-0.7cm}
\begin{center}
\renewcommand{\arraystretch}{1.2}
\begin{tabular}{|c|c|c|c|c|} \hline
Channel \ho\Zo\ra & $\mathcal{L}$ [pb$^{-1}$] & Data & Total bkg & eff [\%]
 \\\hline\hline
\multicolumn{5}{|c|}{\bf\boldmath $192 ~\le~ \sqrts~ \le ~202 $~GeV} \\\hline\hline
{\small{\bb\qq\ ($\mh~=~100$ GeV)}}  & 217.0  & 30 & $28.8 \pm 4.2 $ &  $42.0 $ \\ \hline
{\small{\bb\nn}}  & 212.7  & 10 & $13.9 \pm 1.6 $ &  $46.9 $ \\\hline
{\small{\bb\tautau}}/{\small{\tautau\qq}}
                  & 213.6  & 5 & $5.1 \pm 0.8$ &  $25.8 $ \\\hline
{\small{\bb\ee}}  & 214.1& 3  & $4.1 \pm 0.7$ &  $57.2 $ \\\hline
{\small{\bb\mm}}  & 213.6 & 6  & $3.3 \pm 0.5$  &  $62.5 $ \\\hline
\hline
\multicolumn{5}{|c|}{\bf\boldmath $200 ~\le~ \sqrts ~\le~ 209 $~GeV} \\\hline\hline
{\small{\bb\qq\ ($\mh~=~115$ GeV)}}  & 207.3  & 20 & $17.5 \pm 2.6$ &  $40.0 $ \\ \hline
{\small{\bb\nn}}  & 207.2  &  11 & $8.9 \pm 1.0 $  &  $40.7 $ \\\hline
{\small{\bb\tautau}}/{\small{\tautau\qq}}
                  & 203.6  & 5  & $ 4.5\pm 0.7 $     &  $25.6 $ \\\hline
{\small{\bb\ee}}  & 203.6  & 1  & $ 3.6 \pm 0.7$ &  $52.9  $ \\\hline
{\small{\bb\mm}}  & 203.6  & 4  & $3.4\pm 0.5$  &   $59.2 $ \\\hline
%
\end{tabular}
\caption{\label{tab:smflow}
  The \ho\Zo\ b-tagging channels for data collected in the year 1999 with 
  $192 \le \sqrt{s} \le 202$ GeV, and for data collected in the
  year 2000 with $200 \le \sqrt{s} \le 209$ GeV, respectively.
  The integrated luminosities ($\mathcal{L}$), 
  the numbers of events after the final likelihood or Neural Network cut for 
  the data and the expected background, normalised to the data luminosity are 
  shown.
  The errors on the total background include modeling 
  uncertainties and Monte Carlo statistics.
  The last column shows the detection efficiencies for a Higgs boson with
  \mh~=~100~GeV for the year 1999 data and 
  \mh~=~115~GeV for the year 2000.
  Since the four-jet channel relies on a mass-dependent analysis,
  the numbers quoted in the table are given as an example for 
  \mh = 90~GeV (100) for the year 1999 (2000) data.
  For the four-jet channel, the efficiency is computed only
  for \ho\ra~\bb\ decays, while for the  missing-energy, electron and muon channels
  the efficiency is for all decays of the \ho, assuming SM branching
  fractions.  For the tau channel, the
  efficiency is quoted for the processes \Zo\ho\ra\tautau(\ho\ra all) 
  or \Zo\ho\ra\qq\tautau\ assuming SM branching fractions.
}
\end{center}
\end{table}
\vspace*{-0.4cm}
\hspace{-0.75cm}

\begin{table}
\begin{center}
\renewcommand{\arraystretch}{1.2}
\begin{tabular}{|c|c|c|c|c|} \hline
Channel \ho\Zo\ra & $\mathcal{L}$ [pb$^{-1}$] & Data & Total bkg. &  eff [\%]
 \\\hline\hline
\multicolumn{5}{|c|}{\bf\boldmath $ 192~<~ \sqrts ~<~ 202 $~GeV} \\\hline\hline
{\small{\qq\qq} ($\mh = 90$ GeV)}  & 217.0 & 290 & $290.6 \pm 37.7$ &  $52.0 $ \\ \hline
{\small{\qq\nn}}  & 212.7  & 68 & $70.8 \pm 10.8$ &  $45.2 $ \\\hline
{\small{\qq\tautau}}/{\small{\tautau\qq}}
                  & 213.7  & 1  & $ 4.6 \pm 0.8$ &  $27.1 $ \\\hline
{\small{\qq\ee}}  & 214.1 & 13  & $8.3 \pm 2.5$ & $58.6  $ \\\hline
{\small{\qq\mm}}  & 213.6 & 7  & $7.6 \pm 1.5$  & $64.8 $ \\\hline
\hline
\multicolumn{5}{|c|}{\bf\boldmath $ 200~<~ \sqrts ~<~ 209 $~GeV} \\\hline\hline
{\small{\qq\qq} ($\mh = 100$ GeV)}  & 207.3 & 263 & $235.3 \pm 28.2$ &  $55.0 $ \\ \hline
{\small{\qq\nn}}  & 208.2  & 55 & $62.3 \pm 2.2$ &  $48.4  $ \\\hline
{\small{\qq\tautau}}/{\small{\tautau\qq}}
                  & 205.3  & 2  & $ 4.2 \pm 0.8$ &  $21.0 $ \\\hline
{\small{\qq\ee}}  & 208.2 & 10  & $8.3 \pm 2.5$ & $58.8  $ \\\hline
{\small{\qq\mm}}  & 207.8 & 9  & $7.4 \pm 1.4$  & $62.4 $ \\\hline
\end{tabular}
\caption{\label{tab:flavflow}
  The \ho\Zo\  flavour independent channels for data collected in 1999
  with $192 \le \sqrt{s} \le 202$~GeV and in year 2000 with  $200 \le \sqrt{s} \le 209$~GeV:
  the integrated luminosities ($\mathcal{L}$),
  the numbers of events after the final likelihood or Neural Network cut for 
  the data and the expected background, normalised to the data luminosity.
  The errors on the total background include modeling uncertainties and
  Monte Carlo statistical errors. 
  The last column shows the detection efficiency 
  for a Higgs boson  decaying to quark or gluon pairs with $\mh = 90$~GeV and with $\mh = 100$~GeV
  for data collected in year 1999 and 2000, respectively.
  Since the four-jet channel relies on a mass-dependent analysis, 
  the numbers quoted in the table are given as an example for 
  $\mh = 90$~GeV (100) for the year 1999 (2000) data.
}
\end{center}
\end{table}

\begin{table}
\vspace*{-0.7cm}
\begin{center}
\renewcommand{\arraystretch}{1.2}
\begin{tabular}{|c|c|c|c|c|c|} \hline
Channel \ho\Ao\ra & (\mh,~\mA) [GeV] & $\mathcal{L}$ [pb$^{-1}$] & Data & Total bkg  &eff [\%] \\
\hline
\hline
\multicolumn{6}{|c|}{\bf\boldmath $192 ~\le~ \sqrts~ \le ~209 $~GeV} \\\hline\hline
{\small{\bb\bb }}& (90, 90)  & 424.3  & 22 & $19.9 \pm 2.01 $     &  $49.4 $ \\
\hline
{\small{\bb\tautau\ / \tautau\bb}}& (90, 90)  & 417.2  & 13 & $13.2 \pm 2.02 $  &  $42.5 $ \\ 
\hline
{\small{\bb\bb\bb}}  &(100, 40)    & 424.3  & 22 & 19.9 $\pm$ 2.01 & 59.4\\
\hline
\hline
\multicolumn{6}{|c|}{\bf\boldmath $199 ~\le~ \sqrts~ \le ~209 $~GeV} \\
\hline
\hline
{\small{\bb\bb\nn}}  & (100, 40)   & 207.2 & 19 & 17.2 $\pm$ 2.30  & 66.0\\
\hline
{\small{\bb\bb\qq}}  &(100, 40)                          & 207.3  & 20  & 17.5 $\pm$ 2.6& 31.5\\
\hline
\end{tabular}
\caption{\label{tab:ahflow}
  The \ho\Ao\ channels for data collected in the years 1999 and 2000. 
  The integrated luminosities ($\mathcal{L}$), 
  the numbers of events after the final likelihood or Neural Network cut for 
  the data and the expected background, normalised to the data luminosity.
  The errors on the total background include modeling 
  uncertainties and Monte Carlo statistics.
  The last column shows the detection efficiency for the mass combination
  given in the first two columns.
}
\end{center}
\end{table}

\subsection{External constraints}
\label{subsect:add_exp_constr}

In addition to the combination of the direct search channels, the following external constraints
are applied in every parameter space point considered:

\begin{enumerate} [(a)]
\item 
{A powerful experimental constraint
on extensions of the SM is the determination of the total width
of the \Z\ boson, $\Gamma_{\mathrm{Z}}$, at LEP~\cite{alta}.
Any possible excess width obtained when subtracting
the predicted SM width from the measured value of $\Gamma_{\mathrm{Z}}$,
$\Delta\Gamma_{\mathrm{Z}}$,
can be used to place upper limits on the cross--section of 
\Zo\ decaying, as in the 2HDM, into final states with \ho\ and \Ao\ bosons \cite{mssmpaper172}.
The maximum additional contribution to the total \Z\
width that is compatible with the measured width at 95 $\%$ CL
is $\Delta\Gamma_{\mathrm{Z}} = 6.5$ MeV, obtained from the latest LEP 
combined \Z\ lineshape results \cite{cernew}.
An expected increase of the partial width of the \Z\ is 
evaluated for each scanned parameter space point in the 
2HDM(II); if it is found to exceed the experimental limit,
the point is excluded.
}

\item 
{The decay mode independent search for $\ee\ra\mathrm{S}\Zo$ \cite{pr355},
where S is any scalar particle produced in association with the \Zo\ boson,
provides an upper limit on the 
scaling factor $k$, defined as
$ \sigma_{{\mathrm{S Z}}}=  k\sigma^{\mathrm{SM}}_{\mathrm{H}\mathrm{Z}}$,
where $ \sigma_{{\mathrm{S Z}}} $ is the production cross-section
for a scalar ${\mathrm{S}}$ in association 
with a \Z,
and $\sigma^{\mathrm{SM}}_{\mathrm{H}\mathrm{Z}}$ is the expected SM cross-section for 
${{\mathrm{m_{S}}}}=m_{\mathrm{H}_{SM}}$. This translates into a limit on the
production cross-section in each parameter space 
point of the 2HDM(II) for which $ \sigma_{\mathrm{hZ}} >  k\sigma^{\mathrm{SM}}_{\mathrm{H}\mathrm{Z}}$ at 
95 $\%$ CL.}
\item
{In regions of the 2HDM(II) parameter space 
for which $4 \le \mh\ (\mA) \le 12$ GeV 
a special study was performed in \cite{yuka},
and 95 $\%$ CL limits were obtained on the Yukawa couplings
of \h\ and \Ao\ to down-type fermions. These limits 
are applied as an external constraint in the scan of the 
the 2HDM(II) parameter space described in Section~\ref{section:limits}.
}
\end{enumerate}

The production of any neutral low mass scalar 
particle in association with the \Zo\ was investigated 
in \cite{higgsmall} and, 
for $\mh \le 9.5$ GeV, a mass-dependent upper
limit on the Higgs boson production cross--section
was obtained. This limit was translated in \cite{2hdmpaper} into
an upper limit on the production cross--section
for \mh\ below 9.5 GeV, which was considered as an external 
constraint in combination with the \Z-width constraint (item (a) 
in the previous list).
The decay mode independent
results included as an external constraint (item (b) in the previous list)
provide better exclusion power than the upper limit on Higgs boson production cross--section
obtained in \cite{higgsmall}, increasing the exclusion power 
in the 2HDM(II) parameter space for low \mh\ values.
The external constraint on the Yukawa couplings 
extends the exclusion power to regions of the 2HDM(II) parameter space 
with low values of \mh\ and \mA\
and large \tanb. In general, the external constraints 
applied in the present study improve significantly the results
that were obtained in \cite{2hdmpaper}.

\section{Two Higgs Doublet Model interpretations}
\label{section:limits}

The interpretation of the searches for the neutral Higgs bosons
in the 2HDM(II) is done by scanning the parameter space 
of the model. Every (\mh,~\mA,~\tanb,~$\alpha$) point
determines the production cross--section and 
the branching ratios to different final states. 
The 2HDM(II) parameter space covered by the present study is:

\begin{itemize}  
\item{$1 \le \mh \le 130$ GeV, in steps of 1 GeV}
\item{$3 \le \mA \le 220$ GeV, in steps of 1 GeV; \\
      $300 \le \mA \le 500$ GeV, in steps of 100 GeV; \\
      $0.5 \le \mA \le 2.0$ TeV, in steps of 0.5 TeV}
\item{$0.4 \le \tanb \le 40$, in steps of $1^{\circ}$ in $\beta$, from $\beta = 22^{\circ}$ to
$\beta = 88^{\circ}$ and an additional point corresponding to \tanb$~=~$40} 
\item{$\alpha = -\pi/2,~-\pi/4,~0,~\pi/4$, and $\pi/2$}
\end{itemize}  

The values of $\alpha$ are chosen to extend the analysis 
to the particular cases of maximal and minimal mixing  
in the neutral CP-even sector of the 2HDM(II) ($\alpha = \pm \pi/4$ and $\pm \pi/2$, 
respectively)
and of BR(\h\ra\bb) = 0 ($\alpha = 0$).
The extreme cases $\alpha = \pm \pi/2$
are equivalent in the 2HDM(II) since the mass matrix 
of the CP-even neutral Higgs sector, containing the Higgs 
doublets, becomes diagonal.
In ~\cite{2hdmpaper} models with $1 \le \mh \le 100$ GeV (in steps of 1 GeV),
$5 \le \mA \le 100$ GeV (in steps of 1 GeV, and 
larger steps up to 2 TeV), $0.4 \le \tanb \le 58$ (in steps of $1^{\circ}$ in $\beta$)
and  $\alpha = -\pi/2, ~-3\pi/8, ~-\pi/4, ~-\pi/8$ and 0, were considered. 
The present study is extended to cover models
with positive values of $\alpha$ ($\alpha = \pi/4$ and $\pi/2$), 
which are not allowed in the MSSM-like scenarios. 
Furthermore larger values of \mh\ and \mA\ are explored in this study,
due to the increased sensitivity  of the search channels to high 
\mh\ and \mA\ values: the data analysed have been collected  at larger
centre-of-mass energies and with larger luminosities 
than those in ~\cite{2hdmpaper}. 
For \tanb\ $<$ 0.4, the theoretical predictions become 
unreliable. For \tanb\ $>$ 40  the width of the \Ao\ and \ho\ 
becomes non-negligible.
The decay mode independent study introduced in Section~\ref{subsect:add_exp_constr}.[b] 
is providing exclusion from $\mh \approx 1$ GeV down to 1 KeV, where the limit on the cross-section scaling factor,
$k$, with respect to the Standard Model Higgs-strahlung cross-section is of the order of
0.06~\cite{pr355}. Below $\mA \approx 3$ GeV radiative corrections 
become unstable inducing large fluctuations 
in the calculated cross--sections.

In the present study the other two free parameters of the model, 
$m_{\mathrm{H}}$ and $m_{\mathrm{H^{\pm}}}$, 
are set at values of $m_{\mathrm{H}} = $ 210 GeV
and $m_{\mathrm{H^{\pm}}} = $ 1 TeV,  
above the kinematically accessible region  
at LEP. A scan over values of 
the masses $m_{\mathrm{H}}$ and $m_{\mathrm{H^{\pm}}}$ up to
2 TeV has been performed and no change has been 
observed in the production cross--sections and
branching ratios to final state topologies 
of \ho\ and \Ao, as expected from the theory.
The {\tt HZHA} Monte Carlo generator \cite{hzha}
that includes the 2HDM(II) production cross--sections and 
branching ratios of the Higgs particles
has been used to scan the parameter space.
This generator includes next-to-next-to-leading order QCD
corrections and next-to-leading order electroweak corrections.
The branching ratios and cross--sections obtained were cross--checked 
with the results of another generator ~\cite{jan}
in which QCD corrections are computed
only up to next-to-leading order.
The comparison showed an agreement better than 1 $\%$ between the results of the two
programs.

The results of all the individual search channels
at the studied centre--of--mass energies
are combined statistically to provide 95~\% confidence level (CL) limits,
which are extracted using the method explained in 
Section~\ref{sect:combmet}.
By applying the external constraints discussed in
Section~\ref{subsect:add_exp_constr} additional 
regions of the parameter space are excluded at 95 $\%$ CL.
Although the flavour independent channels supply a unique 
way of investigating parameter space regions where
the branching ratio \ho\ra\bb\ or \Ao\ra\bb\ 
is highly suppressed ({\it{e.g.}}, low $\alpha$
and \tanb\ regions), they have a poor sensitivity with respect to
the b--tagging channels outside these regions.
The use of b--tagging information substantially reduces the background
coming from \WW\ events and improves the sensitivity
of observing Higgs bosons even in regions of the 2HDM(II) parameter 
space where only small branching ratios for \h\ra\bb\ are expected.
The flavour independent and b-tagging searches have
candidate events in common, as is also true for the b-tag 4-jets \h\A\ and \h\Z\
searches. To avoid double--counting 
of candidate events, in each of the parameter space points only the channel 
that provides the better
expected confidence level among the ones that 
have some candidates in common is used for the 
extraction of the limits.

The direct searches for the process \ee\ra\h\Z\  (\ee\ra\h\A)
in the \Z\ data contribute 
mainly in the $\mh \le 50$ GeV ($\mh \le 60$ GeV) region.
For $\sqrt{s}  \ge 189$ GeV, since the flavour independent \h\Z\ and \h\A\  analyses
have been performed in the mass regions $\mh \ge 60$ GeV 
and  $\mh,~\mA \ge 30$ GeV, respectively,
 only b-tagging channels are applied below
these masses.
The flavour independent analyses provide exclusion
for the whole \tanb\ range and for the $\tanb < 1$ regions
for $\alpha=0$  and  $\alpha=\pm \pi/4$, respectively.

\begin{figure}
\centerline{ \epsfig{file=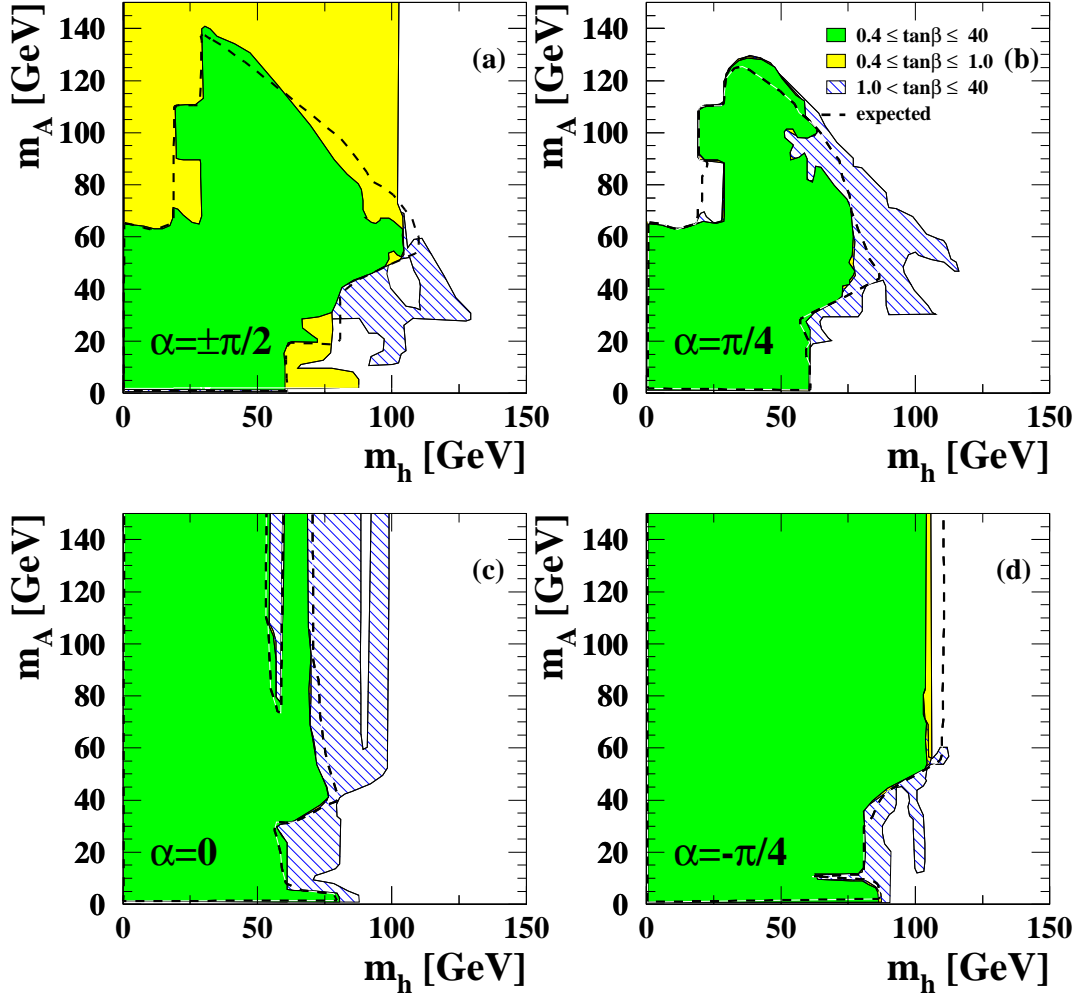,width=\textwidth} }
\caption[] {\label{2hdm1}
Excluded regions in the (\mh,~\mA) plane, (a)--(d), for 
$\alpha=$
$\pm \pi/2$, $\pi/4$, $0$ and $-\pi/4$, respectively,
together with the expected exclusion limits.
A particular (\mh,~\mA,~$\alpha$) point is excluded at  
95\% CL if it is excluded for all scanned values of \tanb.
Three different domains of \tanb\ are shown:
the darker grey region is excluded for all values 0.4 $\le$ \tanb $\le$ 40;
additional enlarged excluded regions are obtained  
constraining 0.4 $\le$ \tanb\ $\le$ 1.0 (lighter grey area) 
or 1.0 $<$ \tanb $\le$ 40 (hatched area).
Expected exclusion limits for 0.4 $\le$ \tanb $\le$ 40 are shown as a dashed line.}
\end{figure}

In Figures \ref{2hdm1}(a--d) the excluded regions in
the (\mh,~\mA) plane are shown for the chosen values of $\alpha$,
together with the calculated expected exclusion limits.  A particular
(\mh, \mA, $\alpha$) point is excluded at 95~\% CL if it is excluded
for all scanned values of \tanb.  Different domains of \tanb\ are
studied and described below: a) $0.4 \le \tanb  \le 40$ and 
b) $0.4 \le \tanb \le 1.0$ or $1.0 < \tanb \le 40$, for which enlarged
excluded regions are obtained.

\vspace{0.3cm}
\hspace{-.6cm}a) $0.4 \le \tanb \le 40$ (darker grey area):

\begin{itemize}
\item {
For all $\alpha$ values, in the region below $\mh \lesssim 10$ GeV
the sensitivity of the channels at $\sqrt{s} \approx \mZ$ is poor.
The exclusion in this region is due to the application of the external
constraints, as explained in section~\ref{subsect:add_exp_constr}, in particular to the Z width 
constraint.
Both the \h\Z\ and \h\A\ production processes contribute to the
natural width of the \Z. 
The exclusion provided by the \h\Z\ process
is valid for any value of \mA. On the other hand,
the exclusion provided by the \h\A\ process is kinematically limited 
to the regions where $\mA + \mh \le  \mZ$. 
The contribution of the \h\Z\
production cross--section to the \Z\ width 
depends on the argument $(\beta - \alpha)$,
and it becomes large
enough for this process alone to provide exclusion 
in different \tanb\ domains for the 
$\alpha$ values considered.}

\item{ The shape of the exclusion plot in Figures \ref{2hdm1}(a) and (b)
for $\mh < 30$ GeV is related to the kinematic constraint on the \h\A\
production in the \Z\ data, which for $\beta - \alpha \approx 0$ or $\pi$ 
is the only allowed process, since the \ho\Zo\ production cross--section 
vanishes. Since the $\alpha$ values are such that the condition
$\beta - \alpha \approx 0$ or $\pi$ is never 
achieved in Figures \ref{2hdm1}(c) and (d), 
the domains with \mh\ $+$ \mA\ $>$ \sqrts\
can be excluded by the \h\Z\ channels. 
For \mh\ $>$ 20-30 GeV, the high energy data 
open a new 
kinematic region and are able to exclude large (\mh, \mA) areas,
as can be seen in Figures~\ref{2hdm1}(a) and (b).
In the same figures the exclusion in the observed rectangular contour 
$20 \lesssim \mh \lesssim 30$ GeV  and $90 \lesssim \mA \lesssim 110$ GeV is due to the recent 
optimisation of the \ee\ra\h\A\ra\bb\bb\ analysis in the same kinematical region
~\cite{mssm020}. The unexcluded region $20~\lesssim~\mh~\lesssim~30$ GeV and
$60~\lesssim~\mA~\lesssim~90$ GeV is due to a small 
excess in the data with respect to the expected background 
of about the same size as the expected signal,
for $\tanb~\approx~8$ in Figure~\ref{2hdm1}(a)
and for $\tanb~\approx~0.7$ in Figure~\ref{2hdm1}(b), respectively.
}

\item{ The (\mh, \mA) points below the semi-diagonal defined by $\mh~\ge~2\mA$,
for which the process \h\ra\A\A\ is kinematically allowed,
can only be excluded by the high energy channels for restricted \tanb\ ranges depending on 
the $\alpha$ values examined. For very low
values of \tanb\ the branching ratio for \A\ra\bb\ vanishes, causing 
unexcluded regions for all values of $\alpha$.
However, these are excluded by the \Z\ data flavour independent 
analyses below $\mh \approx 60$ GeV.
The region for $\mA \le 10$ GeV is difficult to exclude
since the decay \h\ra\A\A\ is usually dominant when kinematically
allowed and the \A\ cannot decay to \bb.
Therefore the searches using b-tagging do not help in this region, while
the sensitivity of the flavour independent \h\Z\ channels is too low to
provide any exclusion.
For $\alpha = 0$ and $\alpha = -\pi/4$, the exclusion in this region is obtained by using 
the dedicated search for the process
\h\Z\ra\A\A\Z\ followed by \A\ra\cc, \tautau, gg and \Z\ra\nn, \mm, \ee.
By applying the Yukawa external constraint large \tanb\ values are excluded.}

\item{The region $55 \le \mh \le 60$ GeV and $\mA \ge 75$ GeV for $\alpha=0$
in Figure ~\ref{2hdm1}(c) is not excluded for $\tanb \le 1$ since for the \Z\ data
the \h\Z\ cross-section becomes too small to exclude it 
and most of the high energy flavour independent channels 
are only efficient for domains in which $\mh \ge 60$ GeV.
}

\item{The largest (\mh, \mA) excluded domain is for $\alpha = - \pi/4$,
where most of the parameter space points accessible at LEP are excluded,
as can be seen in Figure~\ref{2hdm1}(d).}
\end{itemize}

b) $0.4 \le \tanb \le 1.0$ (lighter grey area) 
and  $1.0 < \tanb \le 40$ (hatched area):
\begin{itemize}
\item {In Figure ~\ref{2hdm1}(c), as a consequence of
the variation of the \h\Z\ 
production cross--section with \tanb, as discussed above, 
the $\mh < 10$ GeV and $55 \le \mh \le 60$ GeV, $\mA \ge 75$ GeV regions
are excluded for all values of \mA\ in the $\tanb > 1.0$ domain.
For $\alpha$ = $\pm \pi/2$ 
the $\mh < 30$ GeV region is excluded for all values of \mA\ only in
the $\tanb \le 1.0$ domain.
In the same figure, the region $\mh~\sim~90$ GeV and $\mA \ge 60$ GeV
has an excess of data in the four-jet flavour independent \h\Z\ channel 
at 90 GeV and therefore cannot be excluded even for $\tanb \ge 1$.
}
\item{ At $\alpha=0$ and small values of \tanb\ 
the production cross--section for the process \ee\ra\h\Z\
is highly suppressed. For $\mh > 60$ GeV, constraining
$\tanb~>~1.0$,  larger excluded regions are obtained, 
as can be seen in Figure \ref{2hdm1}(c) (hatched area).
In the same figure, the unexcluded domain $88~\lesssim~\mh~\lesssim~92$ GeV,
$\mA~>~60$ GeV for $\tanb~>~1$ is due to the presence of
candidates in the flavour independent four-jet channel 
in the year 2000 data ~\cite{flavind081}.}
\item{For $\alpha = \pm \pi/2$, Figure \ref{2hdm1}(a),
the region $100 \lesssim \mh \lesssim 110$ GeV and $35 \lesssim \mA \lesssim~50$ GeV
is unexcluded due to the presence of candidate events in the 
\h\A\ra\bb\bb\ and \h\A\ra\bb\tautau\  channels in the high energy 
data ~\cite{mssm020}.}
\end{itemize}

\begin{figure}
\centerline{ \epsfig{file=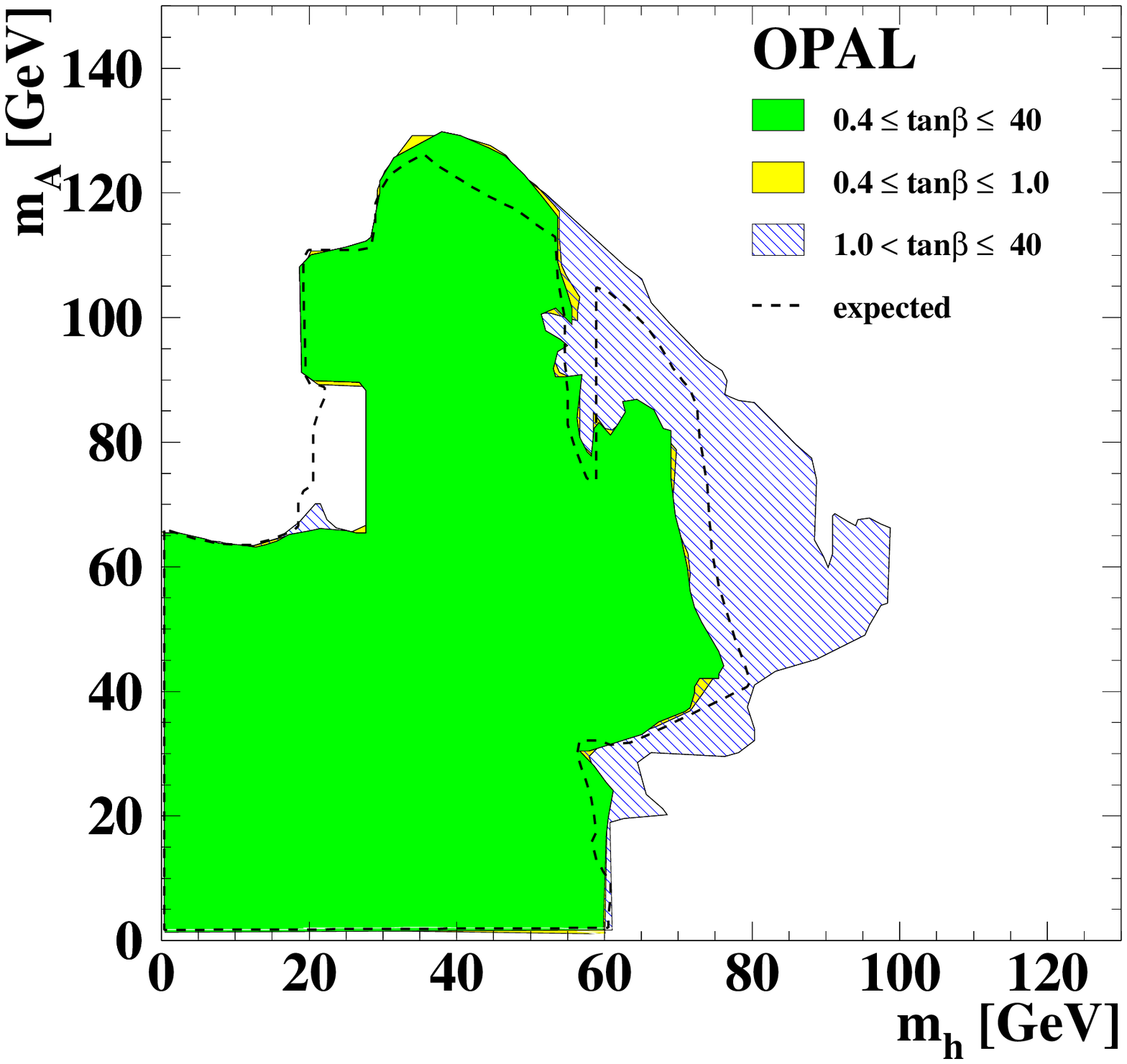,width=\textwidth} }
\caption[]{\label{massmin}
Excluded (\mA,~\mh) region independent of $\alpha$, together with the expected exclusion limit.
A particular (\mA, \mh) point is excluded at  
95\% CL if it is excluded for 0.4 $\le$ \tanb\ $\le$ 40 (darker grey region),
0.4 $\le$ \tanb\ $\le$ 1.0 (lighter grey region) and
1.0 $<$ \tanb\ $\le$ 40 (hatched region) for any $\alpha$.
Expected exclusion limits for 0.4 $\le$ \tanb $\le$ 40 are shown as a dashed line.}
\end{figure}

In Figure \ref{massmin} the excluded regions in the (\mh, \mA) plane
independent of $\alpha$ are given together with the 
expected exclusion limits from MC studies.
A particular (\mh, \mA ) point is excluded at  
95~\% CL if it is excluded for all scanned values of \tanb\ and $\alpha$.
Different domains of \tanb\ are
shown: $0.4~\le~\tanb~\le~40$ (darker grey area), 
$0.4~\le~\tanb~\le~1.0$ (lighter grey area) and 
$1.0~<~\tanb~\le~40$ (hatched area), for which enlarged
excluded regions are obtained.
A rectangular region $1~\lesssim~\mh~\lesssim~55$ GeV for
$3~\lesssim~\mA~\lesssim~63$ GeV is fully excluded at 95~\% CL
independent of $\alpha$ and \tanb.
The scanned $\alpha  >  0$ domain is new with respect to ~\cite{2hdmpaper}
and has the effect of restricting the exclusion for \tanb $~\le~$ 1 
to the kinematically accessible region for the \ho\Ao\ production.

\begin{figure}
\centerline{ \epsfig{file=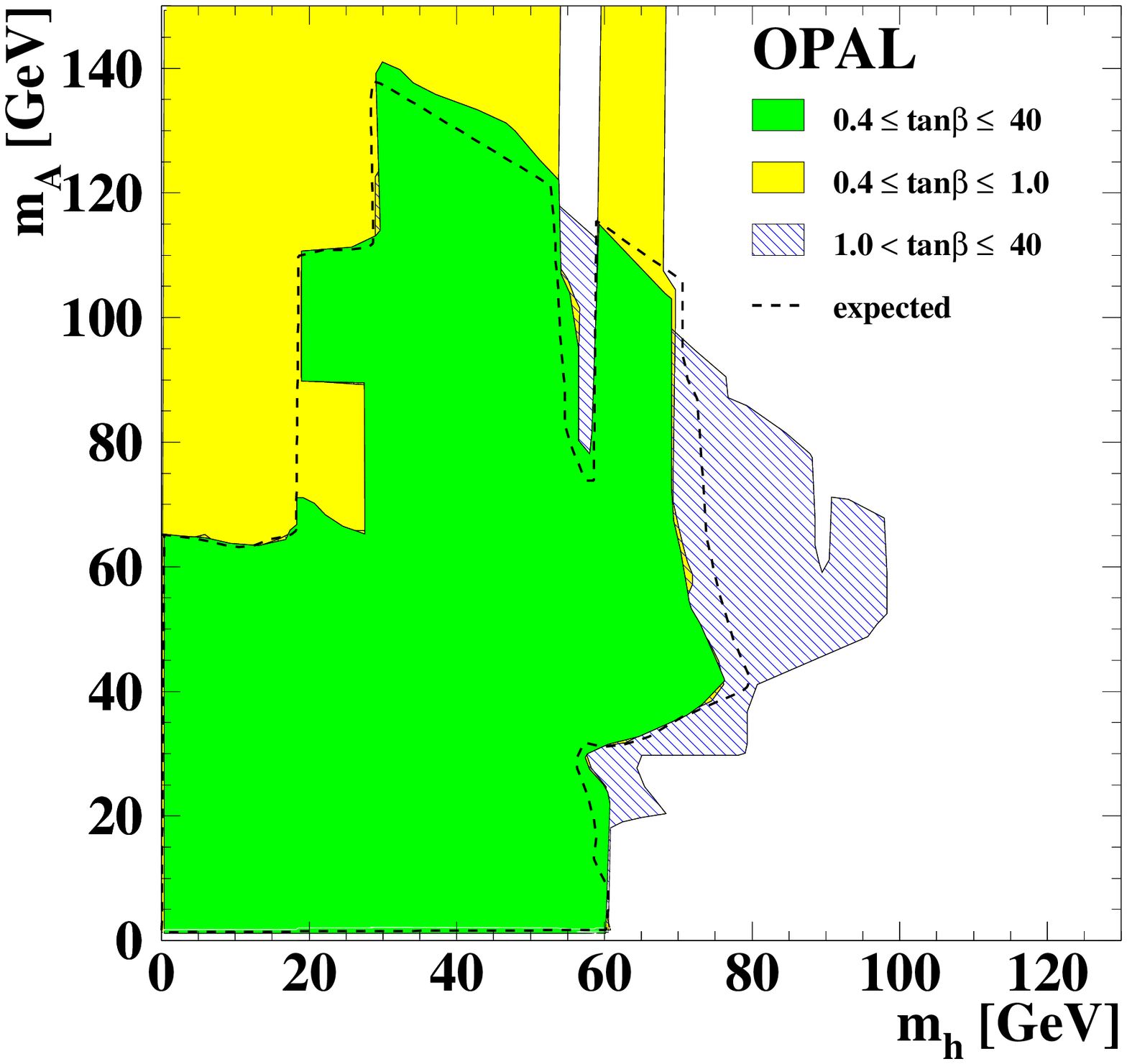,width=\textwidth} }
\caption[]{\label{massneg}
Excluded (\mA,~\mh) region for $-\pi/2~\le~\alpha~\le 0$, together with the expected exclusion limit.
A particular (\mA, \mh) point is excluded at  
95\% CL if it is excluded for 0.4 $\le$ \tanb\ $\le$ 40  (darker grey region),
0.4 $\le$ \tanb\ $\le$ 1.0 (lighter grey region) and
1.0 $<$ \tanb\ $<$ 40  (hatched region). 
Expected exclusion limits for 0.4 $\le$ \tanb $\le$ 40 are shown as a dashed line.}
\end{figure}

For completeness, the excluded regions in the (\mh, \mA) plane
for $\alpha  \le  0$ are given together with the calculated expected 
exclusion limits, is shown in Figure ~\ref{massneg}. 
The present study considerably extends the excluded (\mh, \mA) domain 
for negative values of $\alpha$ when compared with the study 
published in ~\cite{2hdmpaper}. The previously excluded region 
of $1  \lesssim  \mh   \lesssim  44$ GeV
and $12  \lesssim  \mA  \lesssim  56$ GeV is now enlarged to
$1  \lesssim  \mh   \lesssim  55$ GeV
and $3  \lesssim  \mA  \lesssim  63$ GeV,
for all \tanb\ values for negative $\alpha$.

\begin{figure}
\centerline{ \epsfig{file=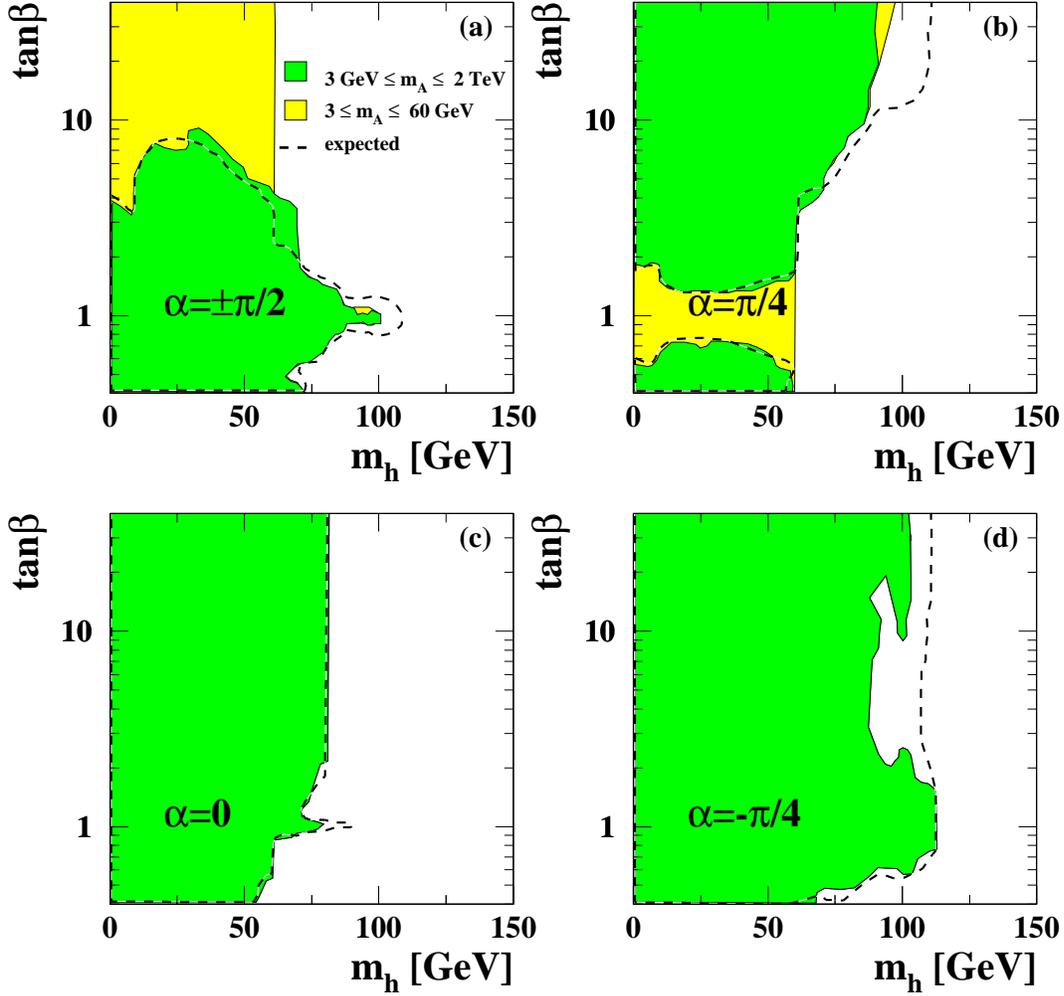,width=\textwidth} }
\caption[]{\label{2hdm2}
Excluded regions in the (\tanb, \mh) plane,
(a)--(d), for $\alpha$=
$\pm \pi/2$, $\pi/4$, $0$ and $-\pi/4$, respectively,
together with the expected exclusion limits.
A particular (\mh, \tanb, $\alpha$) point is excluded at  
95\% CL if it is excluded for all scanned values of \mA.
The two regions shown correspond to the whole domain 3 GeV $\le$ \mA $\le$ 2 TeV (darker grey area)
and a restricted domain for which 3 $\le$ \mA\ $\le$ 60 GeV (lighter grey area). 
The exclusion regions for \mA\ $\le$ 60 GeV entirely contain
the  3 GeV $\le$ \mA $\le$ 2 TeV excluded areas.
Expected exclusion limits are shown for 3 GeV $\le$ \mA $\le$ 2 TeV
(dashed line).
}
\end{figure}

In Figures \ref{2hdm2}(a--d) the excluded regions in the (\tanb, \mh) plane are shown
for the chosen values of $\alpha$,
together with the calculated expected exclusion limits.
A particular (\mh, \tanb, $\alpha$) point is excluded at  
95~\% CL if it is excluded for all scanned values of \mA.
There are two regions shown, the whole domain 3 GeV $ \le $ \mA $ \le $2 TeV (darker grey area)
and a restricted domain for which $3 \le \mA \le 60$ GeV (lighter grey area). The exclusion contours 
for $\mA \le 60$ GeV are larger for all $\alpha$ values, and entirely contain
the  3 GeV $ \le $ \mA $ \le $ 2 TeV excluded areas.

Note that in Figure \ref{2hdm2}(b) the region $\tanb \approx  1$ 
is unexcluded due to the suppression of the \h\Z\
production cross--section as $\beta - \alpha  \approx  0$, while 
relatively low values of \tanb\ can be excluded for $1 \le \mh \le 60$ GeV 
(darker grey area). Restricting the values of \mA\ to be lower than 60 GeV improves the exclusion
since the kinematical limit for \h\A\ production mechanism is never reached. 

\begin{figure}
\centerline{ \epsfig{file=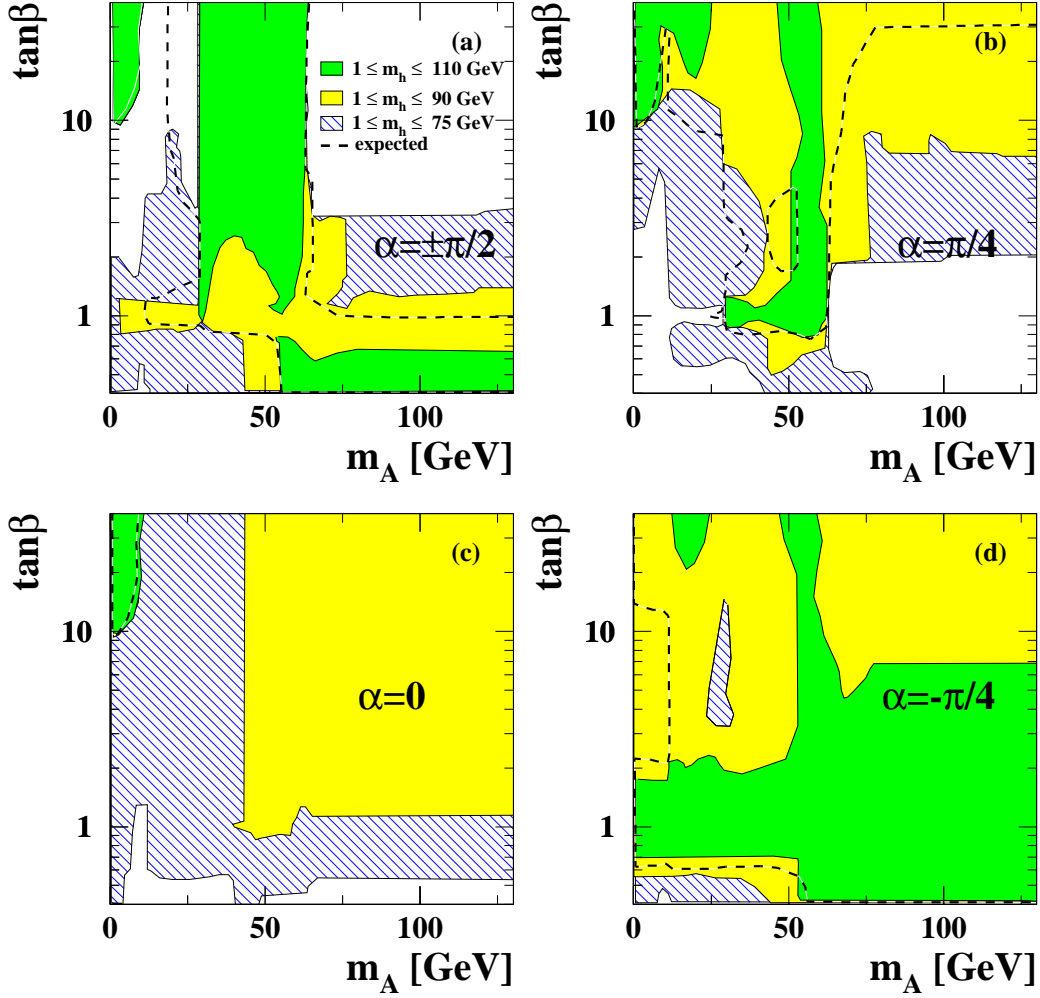,width=\textwidth} }
\caption[]{\label{2hdm3}
Excluded regions in the (\mA, \tanb)
plane, (a)--(d), for $\alpha$=
$\pm \pi/2$, $\pi/4$, $0$ and $-\pi/4$, respectively,
together with the calculated expected exclusion limits. 
A particular (\mA, \tanb, $\alpha$) point is excluded at  
95\% CL if it is excluded for all scanned values of \mh.
The three contours correspond to
1 $\le$ \mh\ $\le$ 110 GeV (darker grey area) and
1 $\le$ \mh\ $\le$ 90 GeV (lighter grey area).
Expected exclusion limits are shown for 1 $\le$ \mh\ $\le$ 110 GeV
(dashed line).
}
\end{figure}

In Figure \ref{2hdm3} the excluded regions in the (\mA, \tanb)
plane are shown for different values of $\alpha$,
together with the expected exclusion limits. 
A particular (\mA, \tanb, $\alpha$) point is excluded at  
95~\% CL if it is excluded for all scanned values of \mh.
There are three regions shown, corresponding to different 
\mh\ domains that are subsets of one another, namely: 
$1 \le \mh \le 110$ GeV (darker grey area),
$1 \le \mh \le 90$ GeV (lighter grey area) and
$1 \le \mh \le 75$ GeV (hatched area).
The lower the \mh\ upper value analysed, the larger the 
excluded (\mA, \tanb) region.

In Figures \ref{2hdm3} (a), (b) and (c) for
$\mA < 10$ GeV and $\tanb > 10$ the excluded domains
are a direct application of the Yukawa external constraint.

Figures \ref{2hdm3} (a) and (b) and Figures \ref{2hdm2} (a) and (b) 
show a similar behaviour in the \tanb\ excluded
domains: for $\alpha  =  \pm \pi/2$ ($\alpha  = \pi/4$) mostly low (large) 
\tanb\ values are excluded.

In Figure \ref{2hdm3} (c) for  
$1 \le \mh \le 90$ GeV the area with
$\mA > 45$ GeV and $\tanb > 1$ is excluded as expected from
the $\tanb > 1$ contour in Figure \ref{2hdm1} (c).
In Figure \ref{2hdm3} (d) the unexcluded region in the 
dark grey contour for small \tanb\
and $\mA < 50$ GeV corresponds to the unexcluded region under the 
semidiagonal in Figure \ref{2hdm1} (d). 
As can be seen in Figure ~\ref{2hdm3} (d) for  
$1 \le \mh \le 90$ GeV, the area 
$24 < \mA < 32$ GeV for $3 < \tanb < 14$ is 
not excluded, but it is excluded  when
$1 \le \mh \le 75$ GeV as inferred from 
Figure \ref{2hdm1} (d).

In Figures \ref{2hdm3} (b) and (d) for  
$1 \le \mh \le 110$ GeV the area
$10 < \mA <  25$ GeV for $\tanb >  20$ is 
excluded thanks to the optimisation  of the \ee\ra\h\A\ra\bb\bb\ analysis in the same kinematical region
~\cite{mssm020}.

\section{Conclusions} 
\label{section:conclusion}

New limits on the \ho\Ao\ pair-production
cross-section are obtained by the application 
of a new flavour independent \h\A\ analysis at the highest LEP energies.
A lower bound at 95 $\%$ CL is extracted along the diagonal
at $\mh \approx \mA \approx 71$ GeV for $c^2=1$ assuming 100 $\%$ 
hadronic decays. The limit
obtained by using the b-tagging analysis~\cite{mssm020}
and assuming 100 $\%$ decays into b-quarks, is at
$\mh \approx \mA \approx 81$ GeV for $c^2=1$.

A general analysis of the 2HDM(II) 
with no CP violation and no extra particles 
besides those of the SM and the five Higgs bosons
has been performed using the \Z, 183 and 189 GeV data
together with the high energy data taken by OPAL in the years
1999 and 2000 at $\sqrt{s} = 192-209$ GeV.
Large areas of the parameter 
space of the model have been scanned. 
In the scanning procedure 
the dependence of the 
production cross--sections and branching ratios
on the angles $\alpha$ and $\beta$,
calculated with next-to-next-to-leading order QCD
corrections and next-to-leading order electroweak corrections, has been considered. 

In addition to the standard OPAL b--tagging analyses, 
flavour independent channels for both the 
Higgs-strahlung process, \ee\ra\ho\Zo, 
and the pair--production process, \ee\ra\ho\Ao,
have been analysed, providing access to regions of parameter 
space in the 2HDM(II) where \ho\ and \Ao\ 
are expected to decay predominantly into up--type light quarks
and gluons ({\it{e.g.}} $\alpha  \approx  0$). 

The 2HDM(II) parameter space scan, for $1 \le \mh \le 130$ GeV,
3 GeV $ \le $ \mA $ \le $ 2 TeV, $-\pi/2 \le \alpha \le  \pi/2$ and $0.4 \le \tanb \le 40$,
leads to large regions being
excluded at the 95~\% CL in the (\mh, \mA) plane
as well as in the (\mh, \tanb) and (\mA, \tanb)
projections. The region $1 \lesssim \mh \lesssim~55$ GeV and
$3 \lesssim \mA \lesssim 63$ GeV is excluded at 95~\% CL
both when restricting $\alpha  \le  0$, as in a MSSM-like scenario,
and in the general 2HDM(II) case, independently of $\alpha$ within the 
scanned parameter space. 

The results obtained by adding the data taken by OPAL in the years
1999 and 2000 substantially enlarge the excluded domains in the 
2HDM(II) parameter space accessible by LEP.

\medskip
\bigskip\bigskip\bigskip
\appendix
\par
\section*{Acknowledgements}
\par
We particularly wish to thank the SL Division for the efficient operation
of the LEP accelerator at all energies
 and for their close cooperation with
our experimental group.  In addition to the support staff at our own
institutions we are pleased to acknowledge the  \\
Department of Energy, USA, \\
National Science Foundation, USA, \\
Particle Physics and Astronomy Research Council, UK, \\
Natural Sciences and Engineering Research Council, Canada, \\
Israel Science Foundation, administered by the Israel
Academy of Science and Humanities, \\
Benoziyo Center for High Energy Physics,\\
Japanese Ministry of Education, Culture, Sports, Science and
Technology (MEXT) and a grant under the MEXT International
Science Research Program,\\
Japanese Society for the Promotion of Science (JSPS),\\
German Israeli Bi-national Science Foundation (GIF), \\
Bundesministerium f\"ur Bildung und Forschung, Germany, \\
National Research Council of Canada, \\
Hungarian Foundation for Scientific Research, OTKA T-038240, 
and T-042864,\\
The NWO/NATO Fund for Scientific Research, the Netherlands.\\
\clearpage
\newpage

\clearpage
\newpage

\clearpage
\newpage

\clearpage
\newpage

\clearpage
\newpage

\clearpage
\newpage

\newpage

\newpage

\newpage

\newpage

\newpage

\include{higgs}
\bibliographystyle{phaip}
\bibliography{higgs}
\end{document}